\documentclass{amsart}

\usepackage{algorithm}
\usepackage{multirow}
\usepackage{graphicx}
\usepackage{subfigure}
\graphicspath{{./figures/}}

\newtheorem{theorem}{Theorem}[section]
\newtheorem{lemma}[theorem]{Lemma}

\theoremstyle{definition}

\theoremstyle{remark}
\newtheorem{remark}[theorem]{Remark}

\numberwithin{equation}{section}

\begin{document}

\title[Unique Diffeomorphic $\pmb{\phi}$ based on $\text{det}\nabla\pmb{\phi}$ and $\nabla\times\pmb{\phi}$]{Recent insights on the Uniqueness Problem of Diffeomorphisms determined by Prescribed\\ Jacobian Determinant and Curl}

\author{Zicong Zhou}
\address{Digital Technology and Innovation Center, Siemens Shanghai Medical Equipment Ltd., Pudong New Area, Shanghai 201318, China}
\email{zicong.zhou.ext@siemens-healthineers.com}
\thanks{The first author is an approved collaborator of NIH Grant R03MH120627.}

\author{Guojun Liao}
\address{Department of Mathematics,
	University of Texas at Arlington, Alrington, Texas 76019, USA}
\email{liao@uta.edu}
\thanks{The second author is partially supported by NIH Grant R03MH120627 as the grand PI.}

\date{Submitted by \today.}



\begin{abstract}
Variational Principle (VP) forms diffeomorphisms with prescribed Jacobian determinant (JD) and curl. Examples demonstrate that, (i) JD alone can not uniquely determine a diffeomorphism without curl; and (ii) the solutions by VP seem to satisfy properties of a Lie group. Hence, it is conjectured that a unique diffeomorphism can be assured by its JD and curl (Uniqueness Conjecture). In this paper, (1) an observation based on VP is derived that a counter example to the Conjecture, if exists, should satisfy a particular property; (2) from the observation, an experimental strategy is formulated to numerically test whether a given diffeomorphism is a valid counter example to the conjecture; (3) a proof of an intermediate step to the conjecture is provided and referred to as the semi-general case, which argues that, given two diffeomorphisms, $\pmb{\phi}$ and $\pmb{\psi}$, if they are close to the identity map, $\pmb{id}$, then $\pmb{\phi}$ is identical $\pmb{\psi}$.
\end{abstract}

\maketitle



\section{Introduction}
The field of construction of diffeomorphisms is a fundamental research area in computational geometry, such as, conformal differential geometry achieved remarkable success in surface differential geometry \cite{Gu}. Whereas, the discussion in this paper focuses on characterizing and controlling meaningful distributions of grid points over a planar domain in 2D or a volumetric 3D domain $\mathrm{\pmb{\Omega}}$, for instance \cite{Grajewski}. This is the problem of adaptive generating non-folding grids. For different demands in various applications, there are distinctive methodologies to each tasks \cite{Liseikin}. But, one common goal for many of approaches is to find a differentiable and invertible transformation $\pmb{T}$, i.e., a diffeomorphism, by controlling its JD, which models local cell-size, such as in \cite{DacMos,Huang}. This idea is also applied to build image registration deformations in \cite{ChenY,Joshi}. 

In this line of research, a grid generation method in \cite{Cai} named the deformation method was built based on \cite{Moser}. It had been furtherer developed, in \cite{Zhou}, to generate Higher Order Element grids. The deformation method constructs $\pmb{T}$ with prescribed JD that is defined by a scalar-valued function $0<f_{o} \in C^1$. The essential step in the deformation method is that to solve a $\bf{divergence-curl}$ system (similar to the constraint equations (\ref{ssdconst})). In that system, the divergence is approximated by $f_{o}-1$ and curl is assigned to $\pmb{0}$ because realizing curl beforehand is another challenging task, where the curl models the local cell-rotation. Consequently, the grids generated by the deformation method are not unique when curl information is lacking, in \cite{Liao}. To overcome this problem, the original VP was proposed in \cite{ChenXi} and studied further in \cite{Zhou}. Its mathematical mechanism is designed to control both JD and curl, where the prescribed curl is defined by a sufficiently smooth vector-valued function $\pmb{g}_{o}$. The original VP works well and accurately in 2D, but in 3D cases, it produces stagnated approximations around the true solution. The limitation leaves the original VP acceptable for some engineering tasks, but once it comes to the areas that require high accuracy such as medical image processing, undesired results are accumulated. Later, in \cite{Zhou2}, the VP is fundamentally revised in its theory and developed with more effective algorithms. Surprisingly, the revised VP not only resolves the inaccuracy in 3D scenario, but also regularizes the solution pool of VP, $\{\pmb{T}\}$, which is a Lie group inside $H^{2}_{0}(\mathrm{\Omega})$. More and more numerical tests are done by the revised VP with nearly ideal results as they are expected. It has become to a prioritized curiosity for us to find out whether a diffeomorphism can be uniquely determined with its JD and curl. That forms the uniqueness conjecture which this paper focuses. Fortunately, so far, the uniqueness issue does not undermine VP to produce well approximated grids and in turn allows unmatched pairs of prescribed JD and curl with discrepancies under certain range. Nonetheless, to the least, we would like to understand better about how well an approximation can be made even if the conjecture is false.

In this paper, the uniqueness of such transformations constructed by VP is studied. It is a fundamental and theoretical work for VP and it is phrased by: $\mathbf{Uniqueness}$ $\mathbf{Conjecture}$ --- given two sufficiently smooth transformations with identical boundaries, $\pmb{\phi}$ and $\pmb{\psi}: \mathrm{\Omega}\rightarrow\mathrm{\Omega}$, $\mathrm{\Omega} \subset \mathbb{R}^3$ such that, 
\begin{equation}\label{UniqueCon1}
	\text{det}\nabla\pmb{\phi}=\text{det}\nabla\pmb{\psi},
\end{equation}  
\begin{equation}\label{UniqueCon2}
	\nabla\times\pmb{\phi}=\nabla\times\pmb{\psi},
\end{equation}  
then, $\pmb{\phi}\equiv\pmb{\psi}$ on $\mathrm{\Omega}$.

The construction of this paper includes: (1) in section 2, VP is reviewed and numerically demonstrated, which points out the effect of curl in determination of a diffeomorphism; (2) in section 3, a remark is made describing that a counter example to the uniqueness conjecture must satisfy a particular property and a experimental strategy is presented to test such (counter) examples; (3) a simple case of the uniqueness problem is analyzed, which is an iterative procedure on the Sobolev's space $H^{2}_{0}(\mathrm{\Omega})$. The $Green's$ formula and the $Poincare's$ inequality are used as the key tools for the argument. It turned out that the general case of uniqueness conjecture is still open.

\section{Variational Principle for Grid Generation}\label{VP}
In \cite{Zhou2}, the revised VP is developed. It is briefly reviewed here with its problem description and some intuitive examples revealing its solutions are of a Lie group. 

Given $\pmb{\phi}_{o}$ on the fixed and bounded domain ($\pmb{\omega}=<x,y,z>\in$) $\mathrm{\Omega} \subset \mathbb{R}^{3}$, let a scalar function $f_o>0$ and a vector-valued function $\pmb{g}_o$ on $\mathrm{\Omega}$ satisfy 
\begin{equation}\label{demmand}
	\int_{\mathrm{\Omega}} f_o(\pmb{\omega})d\pmb{\omega} = |\mathrm{\Omega}| \text{ and } \nabla \cdot  \pmb{g}_o= 0 \text{, respectively.}
\end{equation}
Need to find $\pmb{\phi}=\pmb{\phi}_{\pmb{m}} \circ\pmb{\phi}_{o}$, where $\pmb{\phi}_{\pmb{m}}=\pmb{id}+\pmb{u}$ is an intermediate transformation $\mathrm{\Omega}$ with $\pmb{u} = \pmb{0}$ on $\partial\mathrm{\Omega}$, that minimizes the Sum of Squared Differences, 
\begin{equation}\label{ssd1}
	SSD(\pmb{\phi}) = \frac{1}{2}\int_{\mathrm{\Omega}} [(\text{det}\nabla\pmb{\phi} - f_o)^2+|\nabla \times\pmb{\phi}- \pmb{g}_o|^2] d\pmb{\omega}, \quad
\end{equation}	
\begin{equation}\label{ssdconst}
	\text{ subjects to }
	\left\{
	\begin{aligned}
		&
		\begin{aligned}
			&\nabla \cdot  \pmb{\phi}_{\pmb{m}} = f-1 \\
			&\nabla \times \pmb{\phi}_{\pmb{m}} = \pmb{g}
		\end{aligned}
	\end{aligned}\right.
	\hspace{-0.5cm}
	\Rightarrow
	\mathrm{\Delta} \pmb{\phi}_{\pmb{m}} = \nabla f -\nabla \times\pmb{g}=\pmb{F} (f, \pmb{g}) \text{ in } \mathrm{\Omega}.
\end{equation}	
Please note that (\ref{demmand}) is needed to force $f_{o}$ and $\pmb{g}_{o}$ acting like JD and curl, respectively. More details about the computational schemes and measurements of convergence are referred to \cite{Zhou2}. Here, some intuitive examples are presented in figures for the visualizations of its properties. 

\subsection{$\mathbf{Example}$: Effect of the Curl}	
\begin{figure}[H]
	\vspace{-0.20cm}
	\centering
	\subfigure[$\pmb{R}$]{\includegraphics[width=3cm,height=3cm]{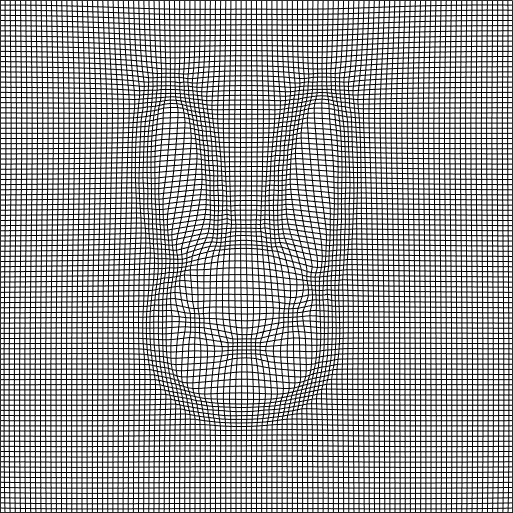}}
	\subfigure[$\pmb{R}_{l}$]{\includegraphics[width=3cm,height=3cm]{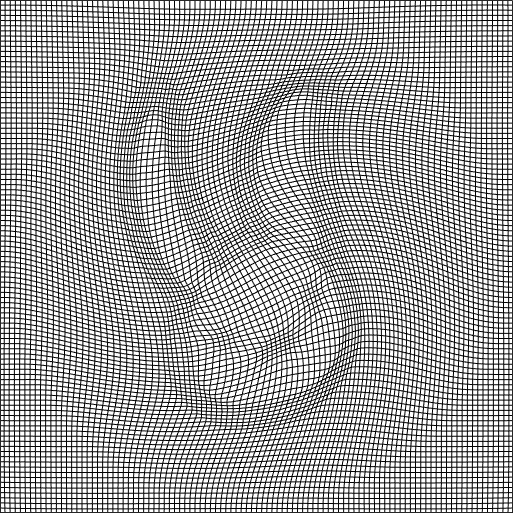}}	
	\subfigure[$\pmb{R}_{r}$]{\includegraphics[width=3cm,height=3cm]{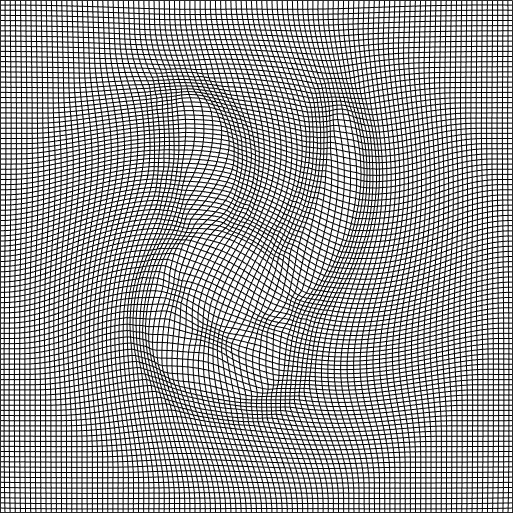}}\\
	\subfigure[Plot of det$\nabla\pmb{R}$]{\includegraphics[width=3cm,height=3cm]{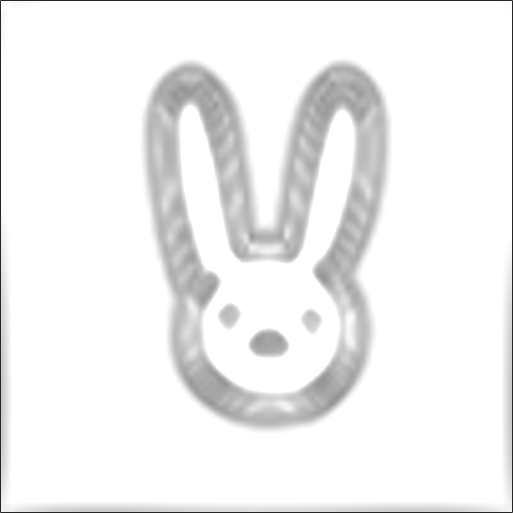}}
	\subfigure[Plot of det$\nabla\pmb{R}_{l}$]{\includegraphics[width=3cm,height=3cm]{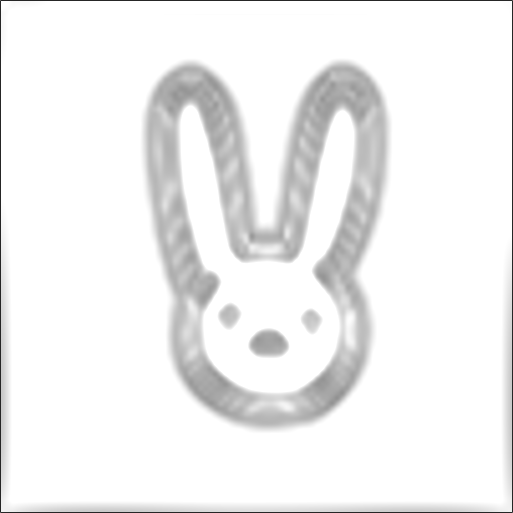}}	
	\subfigure[Plot of det$\nabla\pmb{R}_{r}$]{\includegraphics[width=3cm,height=3cm]{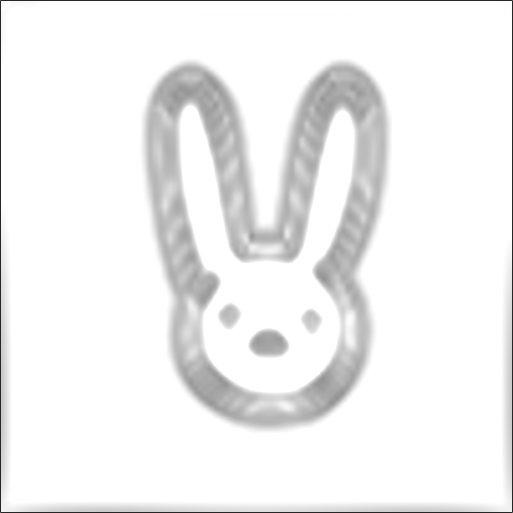}}\\
	\subfigure[Plot of $\nabla\times\pmb{R}$]{\includegraphics[width=3cm,height=3cm]{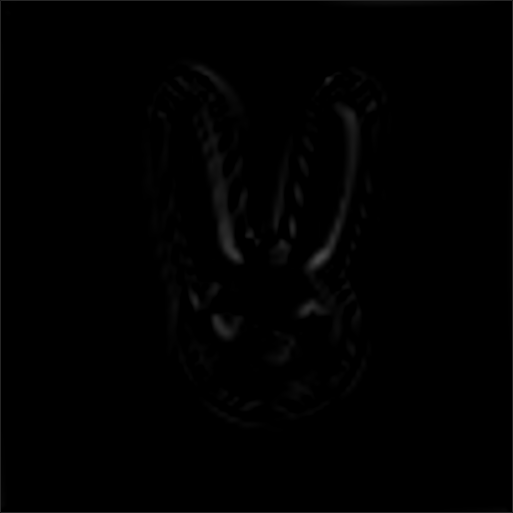}}
	\subfigure[Plot of $\nabla\times\pmb{R}_{l}$]{\includegraphics[width=3cm,height=3cm]{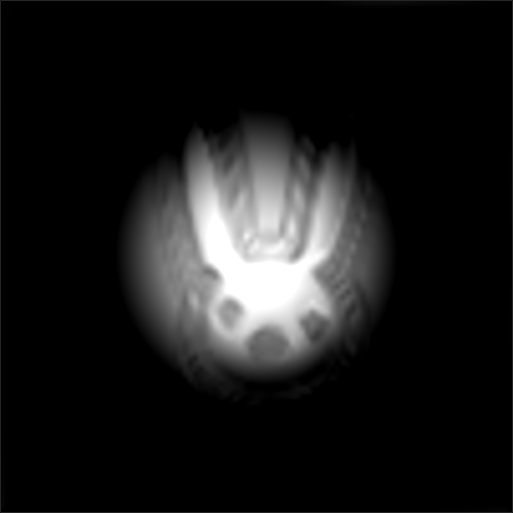}}	
	\subfigure[Plot of $\nabla\times\pmb{R}_{r}$]{\includegraphics[width=3cm,height=3cm]{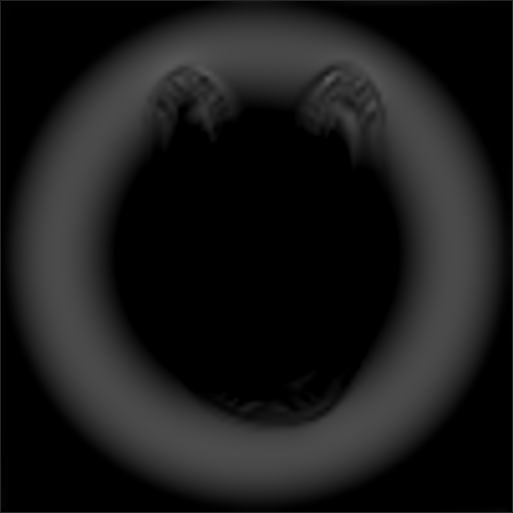}}
	\caption{The Effect of Curl}\label{curly}
	\vspace{-0.3cm}
\end{figure}

Given a grid of a rabbit, $\pmb{R}$, by manipulating curl of $\pmb{R}$, $\nabla\times\pmb{R}$, it is possible to alter $\pmb{R}$ into grids, $\pmb{R}_{l}$ and $\pmb{R}_{r}$, with the same values of JD but different on their curls, namely, $\text{det}\nabla\pmb{R}=\text{det}\nabla\pmb{R}_{l}=\text{det}\nabla\pmb{R}_{r}$ and $\nabla\times\pmb{R}\ne\nabla\times\pmb{R}_{l}\ne\nabla\times\pmb{R}_{r}$. In Fig.~\ref{curly}, such fact is visualized. 

\subsection{$\mathbf{Example}$: Construction of Inverse of a Diffeomorphism}\label{constInv}
\begin{figure}[H]
	\vspace{-0.20cm}
	\centering
	\subfigure[$\pmb{R}$]{\includegraphics[width=3cm,height=3cm]{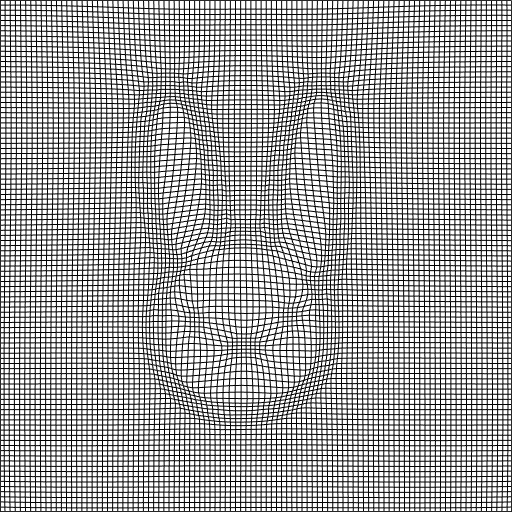}}
	\subfigure[$\pmb{D}$]{\includegraphics[width=3cm,height=3cm]{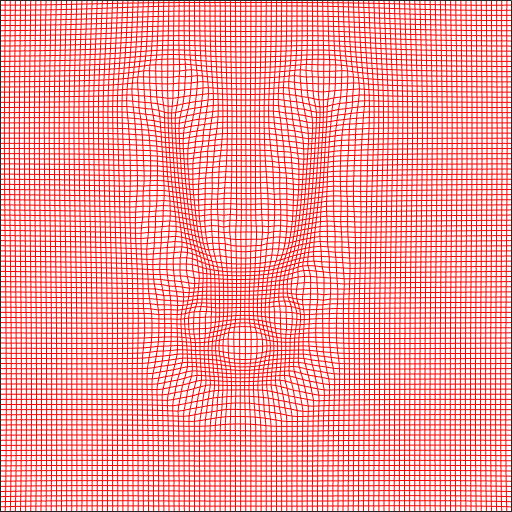}}	
	\subfigure[$\pmb{D}\circ\pmb{R}$ vs $\pmb{id}$]{\includegraphics[width=3cm,height=3cm]{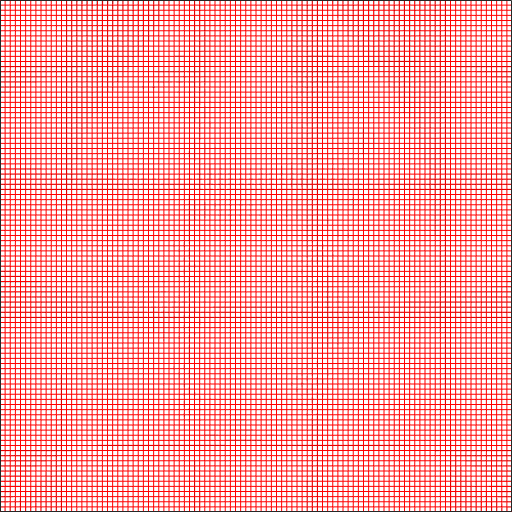}}
	\subfigure[$\pmb{R}\circ\pmb{D}$ vs $\pmb{id}$]{\includegraphics[width=3cm,height=3cm]{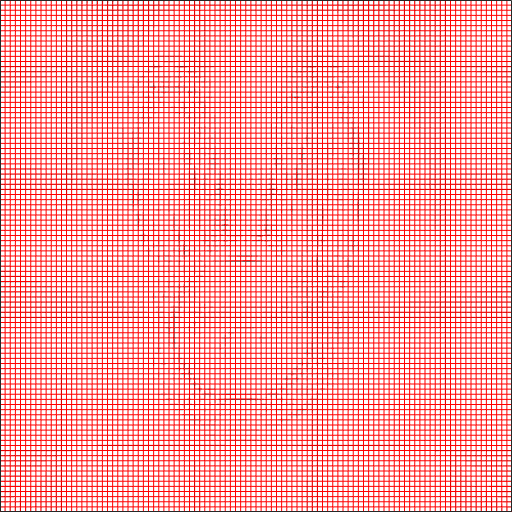}}
	\caption{Construct the Inverse of $\pmb{R}$ by VP}\label{degrid}
	\vspace{-0.3cm}
\end{figure}

This example presents that the VP is capable to construct the inverse transformation of a given diffoemorphism. In Fig.~\ref{degrid}, from the rabbit grid, $\pmb{R}$, its inverse transformation, $\pmb{D}$, is constructed by VP. As Fig.~\ref{degrid} shows in (c) and (d), both $\pmb{R}\circ\pmb{D}$ and $\pmb{D}\circ\pmb{R}$, in red grid lines, are very close to the identity map, $\pmb{id}$, presented by the uniform grid in black grid lines. This is possible because of facts that $\text{det}\nabla\pmb{id}=1=f_{o}$ and $\nabla \times\pmb{id} =\pmb{0}=\pmb{g}_{o}$ hold in the sense of VP. Furthermore, this means VP is effective in constructing diffeomorphisms with prescribed JD and curl.


\subsection{$\mathbf{Example}$: Inverse Consistency and Transitivity}
\begin{figure}[H]
	\vspace{-0.5cm}
	\centering
	\subfigure[$\pmb{D}_{2}: \pmb{R}\rightarrow\pmb{B}$]{\includegraphics[width=2.95cm,height=2.95cm]{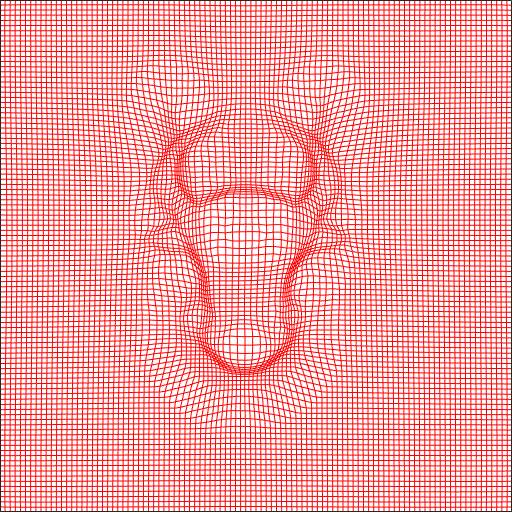}}	
	\subfigure[$\pmb{D}_{2}\circ \pmb{R}$ vs  $\pmb{B}$]{\includegraphics[width=2.95cm,height=2.95cm]{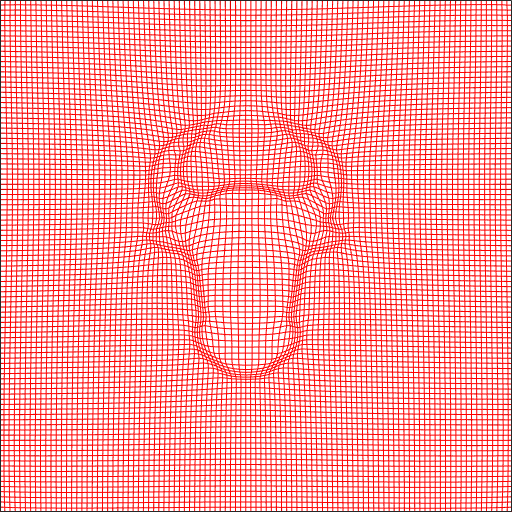}}
	\subfigure[$\pmb{D}_{2}\circ\pmb{D}_{1}$ vs $\pmb{id}$]{\includegraphics[width=2.95cm,height=2.95cm]{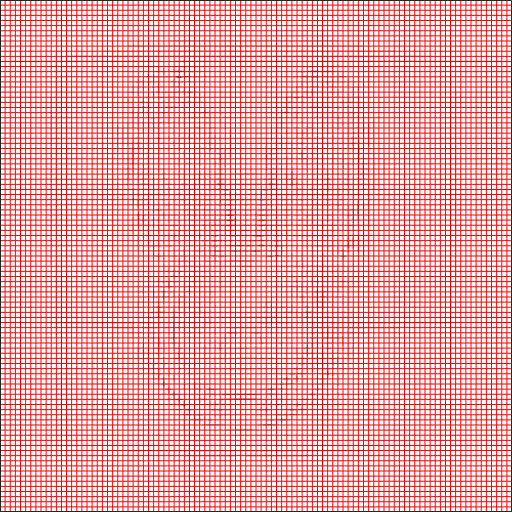}}
	\subfigure[$\pmb{D}_{1}\circ\pmb{D}_{1}$ vs $\pmb{id}$ ]{\includegraphics[width=2.95cm,height=2.95cm]{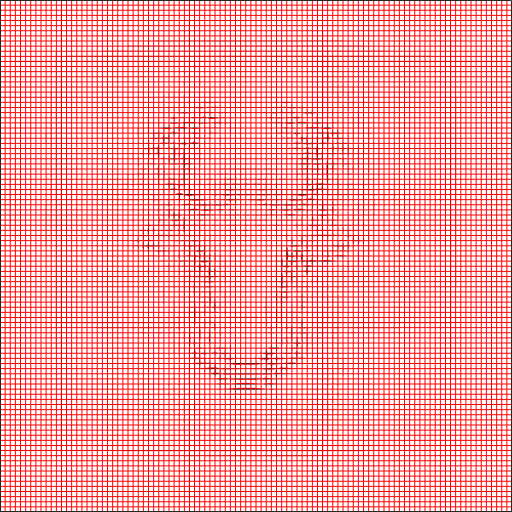}}
	\caption{Inverse Consistency}\label{InvCon}
	\vspace{-0.50cm}
\end{figure}

\begin{figure}[H]
\vspace{-0.5cm}
\centering
	\subfigure[$\pmb{D}_{4}\circ\pmb{C}$ vs $\pmb{R}$ ]{\includegraphics[width=2.95cm,height=2.95cm]{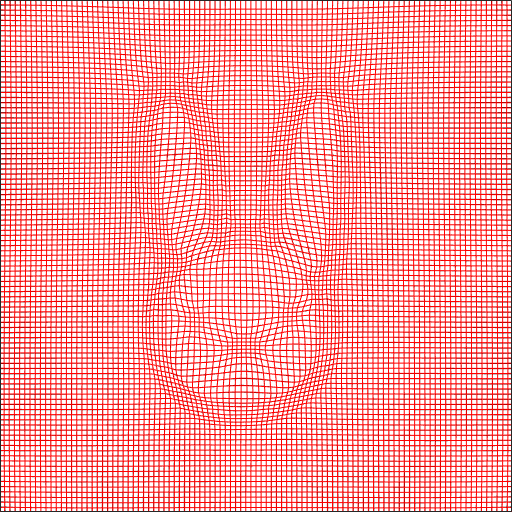}}
	\subfigure[$\pmb{D}_{4\circ3}\equiv\pmb{D}_{4}\circ\pmb{D}_{3}$ ]{\includegraphics[width=2.95cm,height=2.95cm]{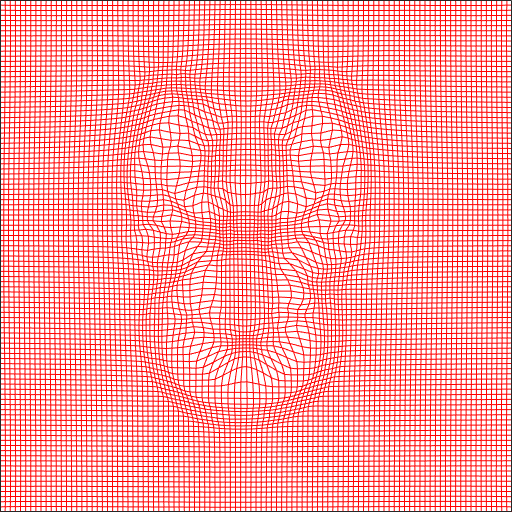}}
	\subfigure[$\pmb{D}_{4\circ3}$ vs $\pmb{D}_{1}$]{\includegraphics[width=2.95cm,height=2.95cm]{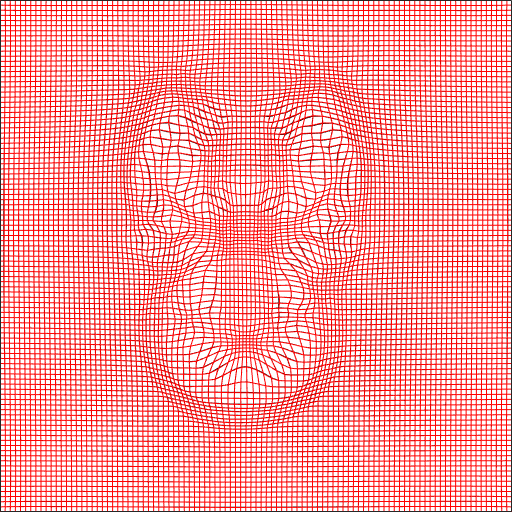}}
	\subfigure[$\pmb{D}_{4\circ3}\circ\pmb{B}$ vs $\pmb{D}_{1}\circ\pmb{B}$]{\includegraphics[width=2.95cm,height=2.95cm]{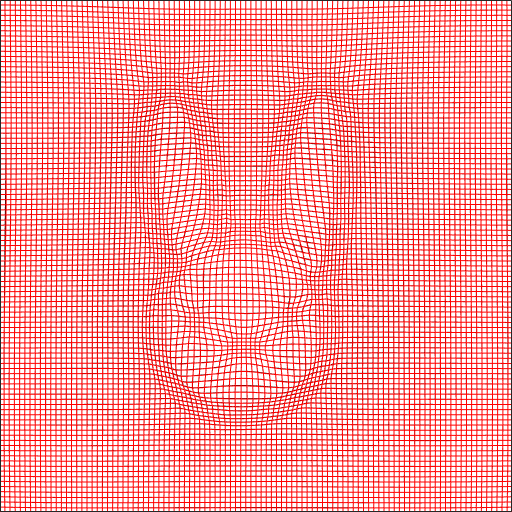}}
\caption{Transitivity}\label{Trans}
\end{figure}

In this example, given another bull grid, $\pmb{B}$, the inverse consistency between $\pmb{B}$ and $\pmb{R}$ are demonstrated in Figure~\ref{InvCon}. Moreover, if one more cat grid, $\pmb{C}$, is inserted between $\pmb{B}$ and $\pmb{R}$, the transitivity of the path $\pmb{B}\rightarrow\pmb{R}$ and the path from $\pmb{B}\rightarrow\pmb{C}$ then $\pmb{C}\rightarrow\pmb{R}$ is demonstrated in Figure~\ref{Trans}. 

Figure~\ref{InvCon}.(d) and (f) are presenting the solutions, $\pmb{D}_{1}\circ \pmb{B}$ and $\pmb{D}_{2}\circ \pmb{R}$, by VP, in red grid lines, are very close to the targets grids, $\pmb{R}$ and $\pmb{B}$, in black grids lines, respectively. Figure~\ref{InvCon}.(g) and (h) are showing that the forward path $\pmb{D}_{1}: \pmb{B}\rightarrow\pmb{R}$ and the backward path $\pmb{D}_{2}: \pmb{R}\rightarrow\pmb{B}$ indeed compensate each other closely to the identity map. $\pmb{D}_{2}\circ\pmb{D}_{1}$ and $\pmb{D}_{1}\circ\pmb{D}_{2}$, in red grid lines, are superimposed on $\pmb{id}$, in black grid lines.
\begin{figure}[H]
	\vspace{-0.5cm}
	\centering
	\subfigure[$\pmb{B}$]{\includegraphics[width=2.95cm,height=2.95cm]{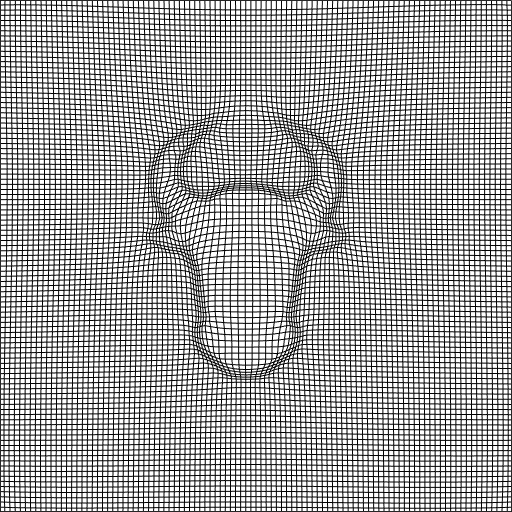}}
	\subfigure[$\pmb{R}$]{\includegraphics[width=2.95cm,height=2.95cm]{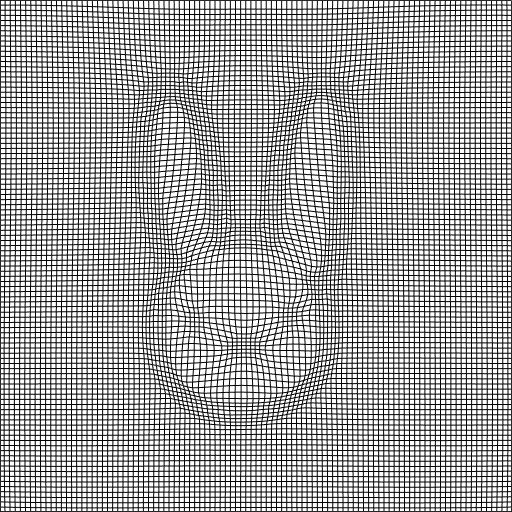}}
	\subfigure[$\pmb{D}_{1}: \pmb{B}\rightarrow\pmb{R}$]{\includegraphics[width=2.95cm,height=2.95cm]{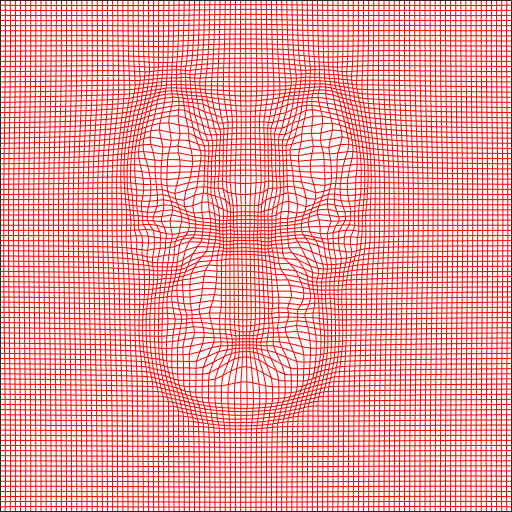}}
	\subfigure[$\pmb{D}_{1}\circ \pmb{B}$ vs  $\pmb{R}$]{\includegraphics[width=2.95cm,height=2.95cm]{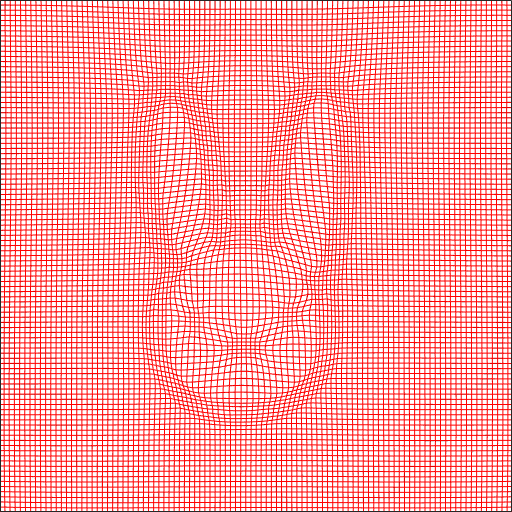}}
	\vspace{-0.5cm}
\end{figure}

Next, in Figure~\ref{Trans}.(b) and (d), are showing the paths of $\pmb{D}_{3}:\pmb{B}\rightarrow\pmb{C}$ and $\pmb{D}_{4}:\pmb{C}\rightarrow\pmb{R}$, whose combination $\pmb{D}_{4\circ3}$ is presented in (f). Figure~\ref{Trans}.(g) indicates that the combined path, $\pmb{D}_{4\circ3}$, in red grid lines, did the same job as $\pmb{D}_{1}$ of Figure~\ref{InvCon}.(c), in black grid lines. Finally, Figure~\ref{Trans}.(h) shows that both $\pmb{D}_{4\circ3}$ and $\pmb{D}_{1}$ leads $\pmb{B}$ to the same end that is very close to $\pmb{R}$. This shows the solutions by VP have the transitivity property.
\begin{figure}[H]
	\vspace{-0.5cm}
	\centering
	\subfigure[$\pmb{C}$]{\includegraphics[width=2.95cm,height=2.95cm]{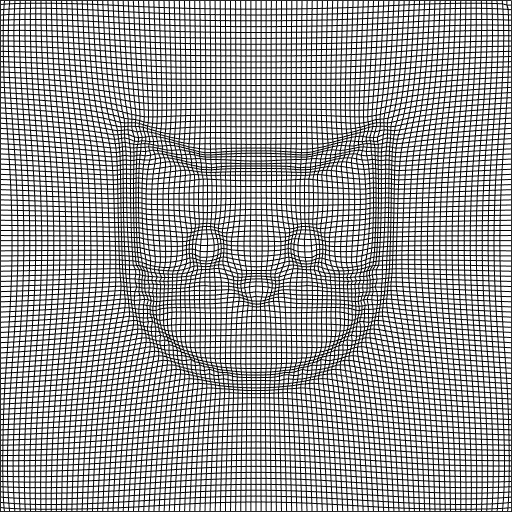}}
	\subfigure[$\pmb{D}_{3}: \pmb{B}\rightarrow\pmb{C}$]{\includegraphics[width=2.95cm,height=2.95cm]{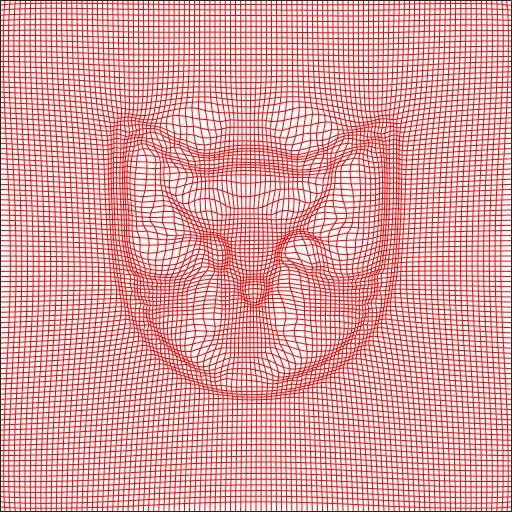}}
	\subfigure[$\pmb{D}_{3}\circ\pmb{B}$ vs $\pmb{C}$]{\includegraphics[width=2.95cm,height=2.95cm]{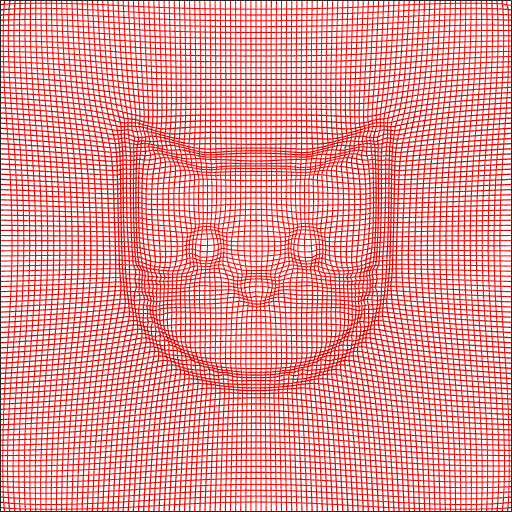}}
	\subfigure[$\pmb{D}_{4}:\pmb{C}\rightarrow\pmb{R}$]{\includegraphics[width=2.95cm,height=2.95cm]{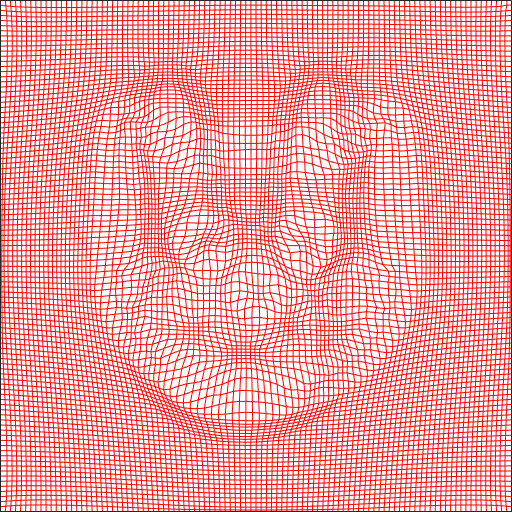}}
	\vspace{-0.5cm}
\end{figure}

\section{An Experimental Strategy to Test Counter Examples}
In this section, a remark of VP is made. We present our observation on this remark to demonstrate a numerical simulation that can be experimented to check whether an example is a counter example to the conjecture. Consider a diffeomorphism $\pmb{\phi}(\pmb{\omega})=\pmb{\omega}+\pmb{u}(\pmb{\omega})=<x+u_1(x,y,z), y+u_2(x,y,z), z+u_3(x,y,z)>$ on $\mathrm{\Omega}$ and the definition of its JD, one may rewrite   
\begin{equation*}
	\text{det}\nabla\pmb{\phi}=\begin{vmatrix}
		1+u_{1x} & u_{2x} & u_{3x}\\ 
		u_{1y} & 1+u_{2y} & u_{3y}\\ 
		u_{1z} & u_{2z} & 1+u_{3z}\notag   
	\end{vmatrix}
	=1+u_{1x}+u_{2y}+u_{3z}
\end{equation*}
\begin{equation*}
	+ u_{1x}u_{2y}u_{3z}+ u_{1z}u_{2x}u_{3y}+ u_{1y}u_{2z}u_{3x}
	- u_{1x}u_{2z}u_{3y}- u_{1y}u_{2x}u_{3z}- u_{1z}u_{2y}u_{3x}
\end{equation*}
\begin{equation*}
	+ u_{1x}u_{2y}+ u_{1x}u_{3z}+ u_{2y}u_{3z}
	- u_{1y}u_{2x}- u_{1z}u_{3x}- u_{2z}u_{3y}
\end{equation*}
\begin{equation}\label{Jd_phi}
	\Rightarrow \text{det}\nabla\pmb{\phi}=1+\nabla\cdot\pmb{u}+\text{det}\nabla\pmb{u}+\text{Tail}(\pmb{
		u}),
\end{equation}
where, we denote $\text{Tail}(\pmb{
		u})= u_{1x}u_{2y}+ u_{1x}u_{3z}+ u_{2y}u_{3z}
	- u_{1y}u_{2x}- u_{1z}u_{3x}- u_{2z}u_{3y}.$
From $SSD$ in (\ref{ssd1}), by prescribing $f_o=1$ and $\pmb{g}_o=\pmb{0}$, $\pmb{\phi}$ can be constructed close to $\pmb{id}$, which is similar to the example presented in section \ref{constInv}. If the uniqueness conjecture were true, then the construction will produce exactly $\pmb{\phi}=\pmb{id}$, otherwise, one should not assume that $\pmb{\phi}$ is necessarily identical to $\pmb{id}$. Suppose the construction of a diffeomorphism is done completely and effectively by VP, that means, 
\begin{equation*}\label{lala1}
	\text{det}\nabla\pmb{\phi}=f_o=1
\text{ and }\nabla\times\pmb{\phi}=\pmb{g}_o=\pmb{0}
\end{equation*}
are attained. This leads to
\begin{equation*}\label{lala2}
	\left\{
	\begin{aligned}
		0=&\text{det}\nabla\pmb{\phi}-1=\nabla\cdot\pmb{u}+\text{det}\nabla\pmb{u}+\text{Tail}(\pmb{u}),\\
		0=&\nabla\times\pmb{\phi}-\pmb{0}=\nabla\times\pmb{u}.
	\end{aligned}\right.
\end{equation*}
Thus, a perspective of $\pmb{u}$ of the uniqueness conjecture can be formed. Let us define 
\begin{equation}\label{Lu}
L(\pmb{u})= L_{1}(\pmb{u})+L_{2}(\pmb{u})=\frac{1}{2}\int_{\mathrm{\Omega}} (\nabla\cdot\pmb{u}+\text{det}\nabla\pmb{u}+\text{Tail}(\pmb{u}))^2d\pmb{\omega}+\frac{1}{2}\int_{\mathrm{\Omega}}|\nabla \times\pmb{u}|^2 d\pmb{\omega}.
\end{equation}
Please note, that $L_{1}(u)$ is under a mathematical mechanism that each terms could potentially compensate each other in general without a necessity for all terms being $0$. Minimizing $L(\pmb{u})$ is somewhat like ``getting rid of" the effects of both JD and curl of $\pmb{\phi}$ toward the displacement vector field $\pmb{u}$. 
\begin{remark}\label{rmk}
	The question about to the conjecture should be phrased as, for any $\pmb{\phi}=\pmb{id}+\pmb{u}$, $\pmb{u}$ vanishes on $\partial \mathrm{\Omega}$, whether 
	\begin{equation*}
		\left\{
		\begin{aligned}
			&\nabla\cdot\pmb{u}+\text{det}\nabla\pmb{u}+\text{Tail}(\pmb{u})=0,\\
			&\nabla\times\pmb{u}=\pmb{0},
		\end{aligned}\right.
		\quad\text{ implies } \quad \pmb{u}=\pmb{0}?
	\end{equation*}
\end{remark}
Therefore, if $\pmb{\phi}$ is a counter example to the conjecture, then, by minimizing $L(\pmb{u})$ in (\ref{Lu}), non-zero $\pmb{u}$ with some significant magnitude will be produced, i.e., such $\pmb{u}$ cannot be continuously shrunk any closer to $\pmb{0}$. 

In the following examples, we numerically tested, in seeking such a counter example, but none has been found so far. Please see section \ref{app1} Appendix for (1) the derivation of the variational gradient of $L(\pmb{u})$ in (\ref{Lu}), under the same constraint as (\ref{ssdconst}), with respect to the control function $\pmb{F}$ and (2) the algorithm summarizing pseudo-code of a gradient-based computational scheme for our tests. In the context of Algorithm~\ref{alg}, given $\pmb{\phi}_{o}=\pmb{id}+\pmb{u}_{o}$, $L(\pmb{u}_{initial}=\pmb{u}_{o})$ is minimized to output $\pmb{u}$. Define $ratio=\frac{L(\pmb{u})}{L(\pmb{u}_{o})}$, by setting $ratio$-tolerance to $10^{-2n}$, Algorithm~\ref{alg} is able to achieve $\left\|\pmb{u}\right\|_{L^{2}}\approx O(10^{-n})$ with proportional iterations allowed. This is confirmed in the following tables. The major computational cost occurs at steps of solving $Poisson$ Equations. In our implementation, a Fast Fourier Transformation $Poisson$ solver is used.

\subsection{$\mathbf{Example}$: Minimize $L(\pmb{u})$ in 2D}
\begin{figure}[H]
	\vspace{-0.5cm}
	\centering
	\subfigure[$\pmb{\phi}_{o}$]{\includegraphics[width=2.95cm,height=2.95cm]{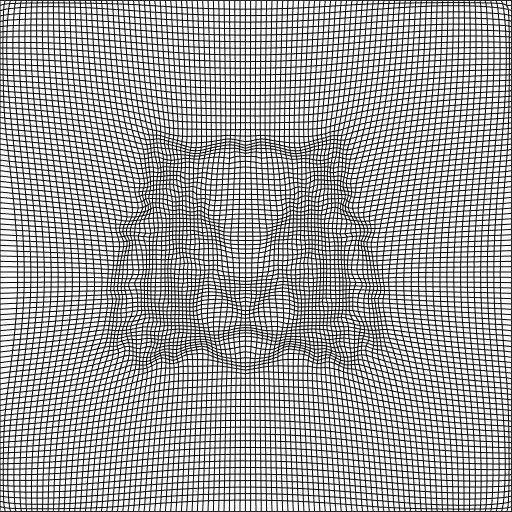}}
	\subfigure[$\pmb{u}_{o}$]{\includegraphics[width=2.95cm,height=2.95cm]{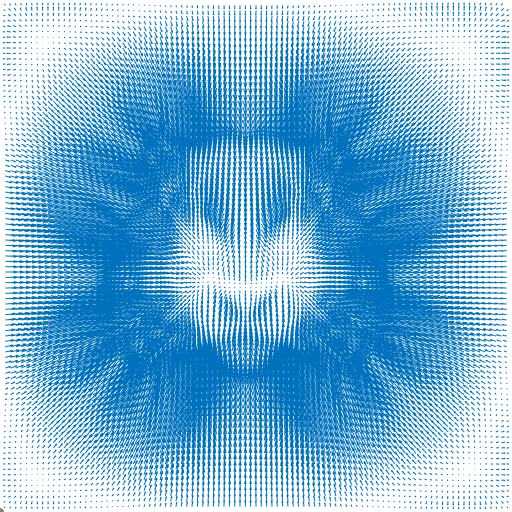}}
	\subfigure[$\pmb{u}$]{\includegraphics[width=2.95cm,height=2.95cm]{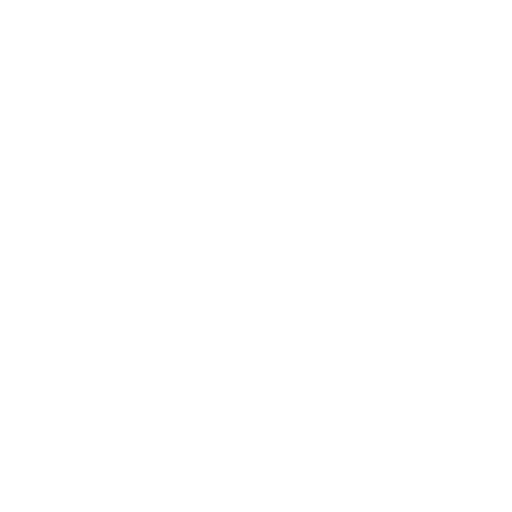}}
	\subfigure[$\pmb{id}+\pmb{u}$ vs $\pmb{id}$]{\includegraphics[width=2.95cm,height=2.95cm]{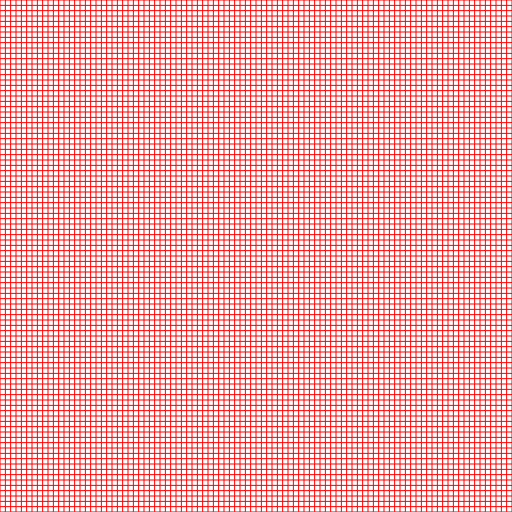}}
	\caption{Shrinking JD and Curl of $\pmb{u}$ in 2D}\label{detiger}
\end{figure}

In Figure~\ref{detiger}(b), the displacement vector field $\pmb{u}_{o}$ has been reduced to (c) $\pmb{u}$ by Algorithm~\ref{alg}. As it can be seen, there is barely any visible vectors. In Figure~\ref{detiger}(d), the output $\pmb{u}$ is added to $\pmb{id}$, in red grid lines, and superimposed on $\pmb{id}$, in black grid lines. This shows that $\pmb{id}+\pmb{u}$ is so close to $\pmb{id}$. In fact, Table~\ref{tb1} recorded the resulting size of $|| \pmb{u}||_{2}$ is only $3.3085*10^{-6}$, with a preset  $ratio$-tolerance$=10^{-12}$ when running Algorithm~\ref{alg}.

\begin{figure}[H]
	\vspace{-0.5cm}
	\centering
	\subfigure[$\hat{\pmb{\phi}}_{o}$ ]{\includegraphics[width=2.95cm,height=2.95cm]{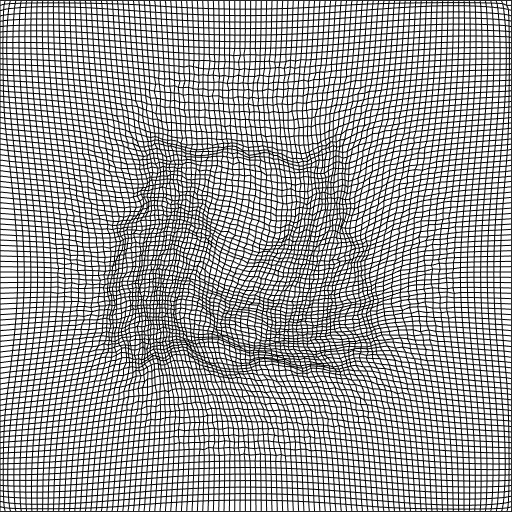}}
	\subfigure[$\hat{\pmb{u}}_{o}$ ]{\includegraphics[width=2.95cm,height=2.95cm]{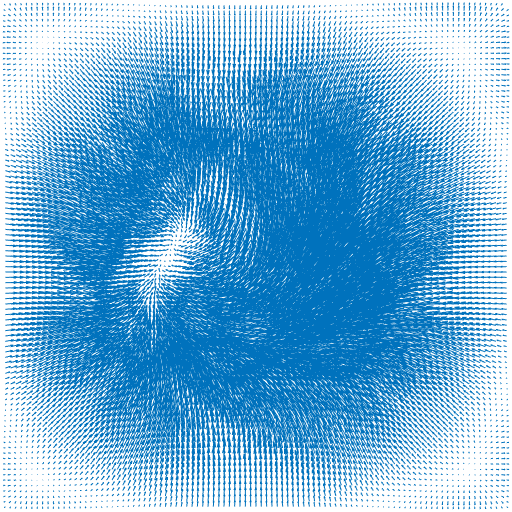}}
	\subfigure[$\hat{\pmb{u}}$]{\includegraphics[width=2.95cm,height=2.95cm]{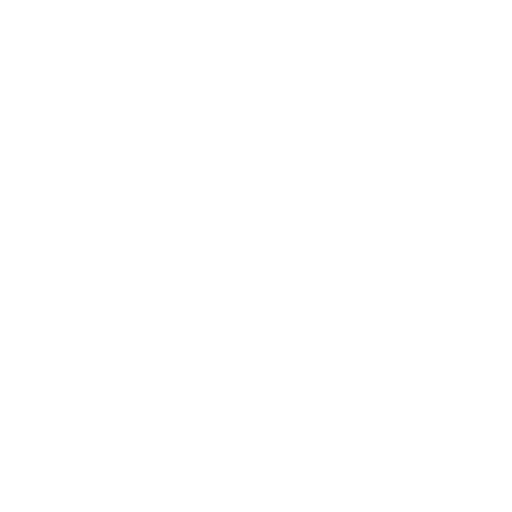}}
	\subfigure[$\pmb{id}+\hat{\pmb{u}}$ vs $\pmb{id}$]{\includegraphics[width=2.95cm,height=2.95cm]{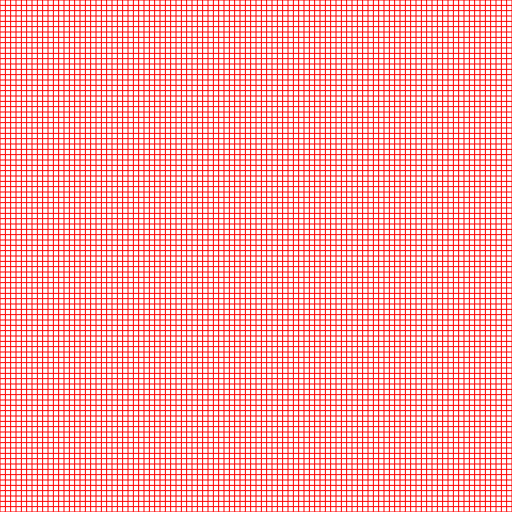}}
	\caption{Shrinking JD and Curl of $\hat{\pmb{u}}$ --- a noised $\pmb{u}$ in 2D}\label{detigern}
\end{figure}

	In Figure~\ref{detigern}, a similar example to Figure~\ref{detiger} is presented with adding Gaussian noise to $\pmb{u}_{o}$ to get $\hat{\pmb{u}}_{o}$ before running Algorithm~\ref{alg}. It shows that the minimization of $L(\pmb{u})$ is not affected by the added noise.
	
\begin{table}[h]
	\begin{center}
		\begin{tabular}{|c|c|c|c|c|}
			\hline	
			\multirow{2}{*}{Solution} & \multirow{2}{*}{$ratio$} & \multicolumn{3}{c|}{max value in} \\  \cline{3-5}
			&    &$|$det$\nabla\pmb{u}^{*}|$ & $||\nabla\times\pmb{u}^{*}||_{2}$ &  $|| \pmb{u}^{*}||_{2}$ \\ \hline
			{\tt $\pmb{u}^{*}=\pmb{u}$} & $9.9893*10^{-13}$& $1.9039*10^{-6}$ &$1.3596*10^{-6}$ &  $3.3085*10^{-6}$ \\
			{\tt$\pmb{u}^{*}=\hat{\pmb{u}}$} & $9.9840*10^{-13}$& $1.3136*10^{-6}$ &$1.3486*10^{-6}$ &  $2.0153*10^{-6}$ \\
			\hline
		\end{tabular}
		\caption{Performance of Figure~\ref{detiger} and Figure~\ref{detigern}}\label{tb1}
	\end{center}
\end{table}



In the Table~\ref{tb2}, it shows a statistical records of 64 repeats of the similar experiment done in Figure~\ref{detigern}, thanks to the accessible Gaussian noise added to $\pmb{u}_{o}$. However, this time, $ratio$-tolerance is dropped to $10^{-14}$, so the resulting sizes of $\left\|\pmb{u}\right\|_{L^{2}}$ is about $O(10^{-7})$. As expected, Algorithm~\ref{alg} performs well and is able to achieve mean size of $\left\|\pmb{u}\right\|_{L^{2}}\approx O(10^{-7})$ with variance about the size $O(10^{-12})$. So, we are looking at a spike like normal distribution, which also indicates that Algorithm~\ref{alg} does not get affected by the added noised.

\begin{table}[h]
	\begin{center}
		\begin{tabular}{|c|c|c|c|}
			\hline	
			\multirow{2}{*}{Total 64 tests} &\multicolumn{3}{|c|}{ max values of } \\  \cline{2-4}
			&  $|$det$\nabla\pmb{u}|$ & $||\nabla\times\pmb{u}||_{2}$ &  $|| \pmb{u}||_{2}$ \\ \hline
			mean & $4.4002*10^{-7}$ &$3.2690*10^{-7}$ &  $6.5365*10^{-7}$ \\
			variance & $2.4486*10^{-15}$ &$8.1358*10^{-16}$ &  $5.0800*10^{-14}$ \\
			\hline
		\end{tabular}
		\caption{Performance of 64 noised tests similar to  Figure~\ref{detigern}}\label{tb2}
	\end{center}
\end{table}

%

\subsection{$\mathbf{Example}$: Minimize $L(\pmb{u})$ in 3D}
\begin{figure}[H]
	\vspace{-0.5cm}
	\centering
	\subfigure[$\pmb{\phi}=\pmb{id}+\pmb{u}$ vs $\pmb{id}$ ]{\includegraphics[width=2.95cm,height=2.95cm]{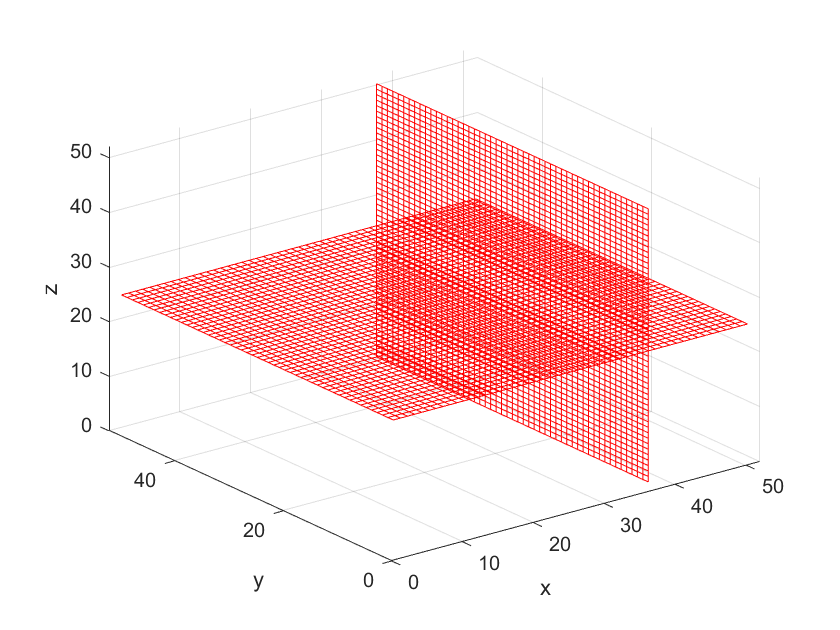}}
	\subfigure[$\pmb{\phi}$ on xz-plain ]{\includegraphics[width=2.95cm,height=2.95cm]{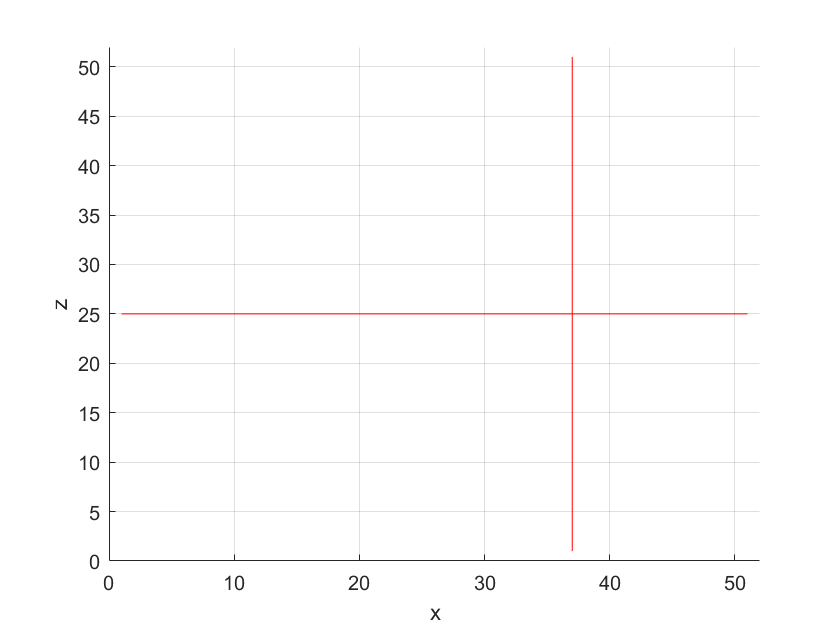}}
	\subfigure[$\pmb{u}$]{\includegraphics[width=2.95cm,height=2.95cm]{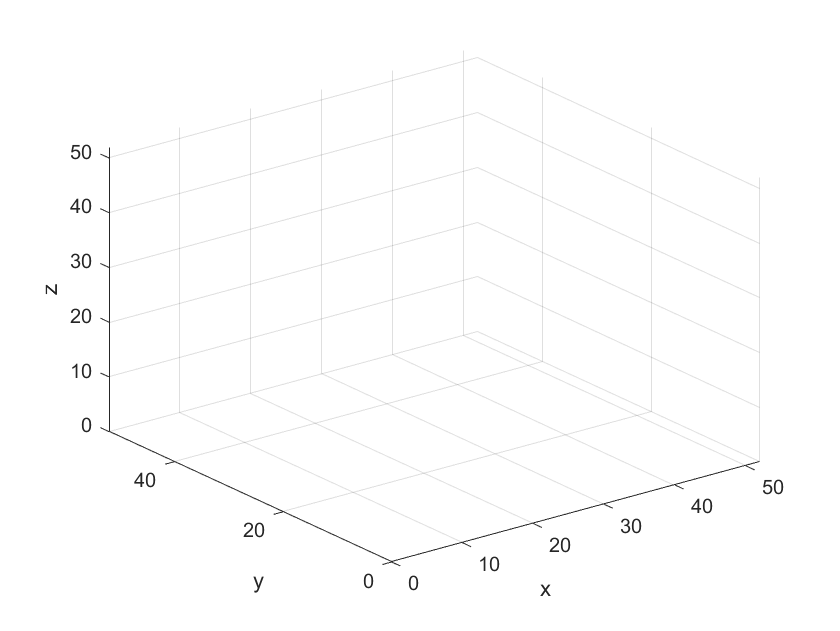}}
	\subfigure[$\pmb{u}$ on xz-plain]{\includegraphics[width=2.95cm,height=2.95cm]{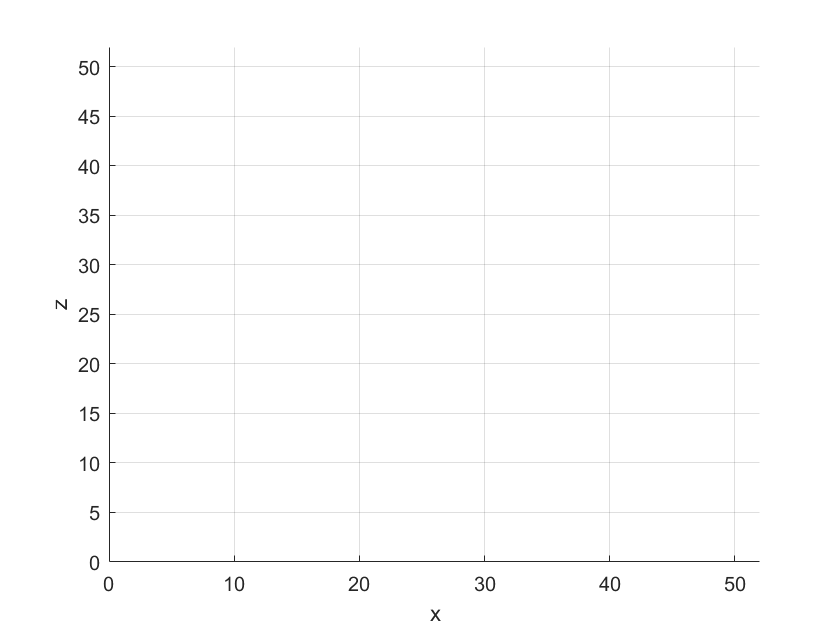}}
	\caption{Shrinking JD and Curl of $\pmb{u}$ in 3D}\label{dewave}
	\vspace{-0.5cm}
\end{figure}

For clean visualizations, the 3D presentations of grids, displacement vector fields and their projections onto the $xz$-plain are only plotted with the 25-th frame on $z$-axis and the 37-th frame on $x$-axis. In Figure~\ref{dewave}(a,b), $\pmb{\phi_{o}}$ is built by multiple times of applying cut-off rotation, displacement, skewing, etc, to $\pmb{id}$. Then, $\pmb{u}_{o}=\pmb{\phi_{o}}-\pmb{id}$ is acquired and showed in Figure~\ref{dewave}(c,d). This is a similar example to the previous one other than being in 3D scenario. Minimize $L(\pmb{u})$ by running Algorithm~\ref{alg} with preset $ratio$-tolerance$=10^{-12}$, size of $||\pmb{u}||_{2}$ gets to $3.002*10^{-6}$.

\begin{figure}[H]
	\vspace{-0.5cm}
	\centering
	\subfigure[$\pmb{\phi}_{o}$]{\includegraphics[width=2.95cm,height=2.95cm]{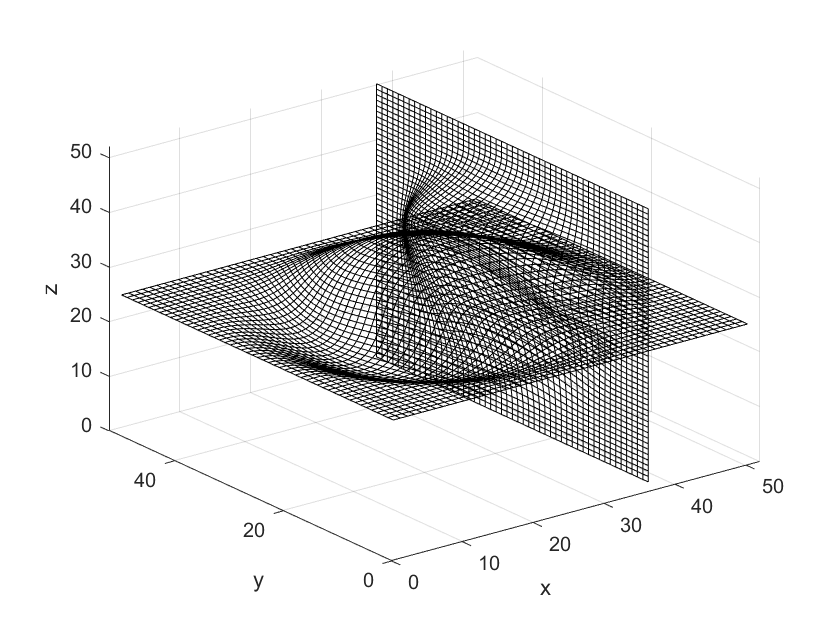}}
	\subfigure[$\pmb{\phi}_{o}$ on xz-plain]{\includegraphics[width=2.95cm,height=2.95cm]{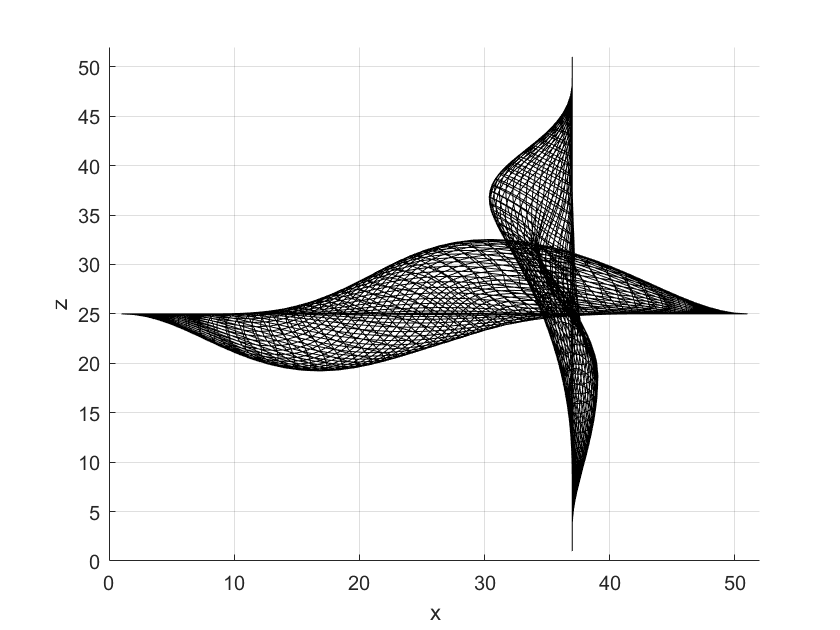}}
	\subfigure[$\pmb{u}_{o}$]{\includegraphics[width=2.95cm,height=2.95cm]{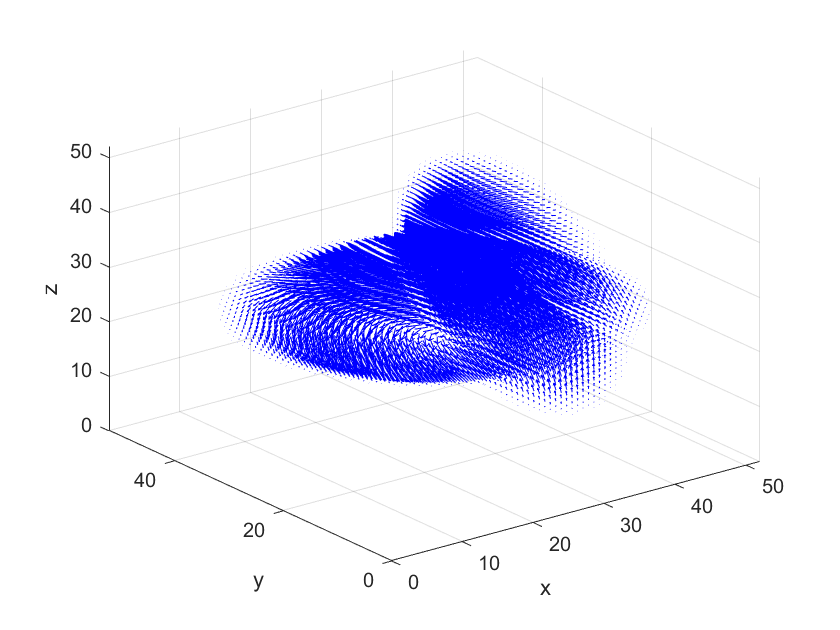}}
	\subfigure[$\pmb{u}_{o}$ on xz-plain]{\includegraphics[width=2.95cm,height=2.95cm]{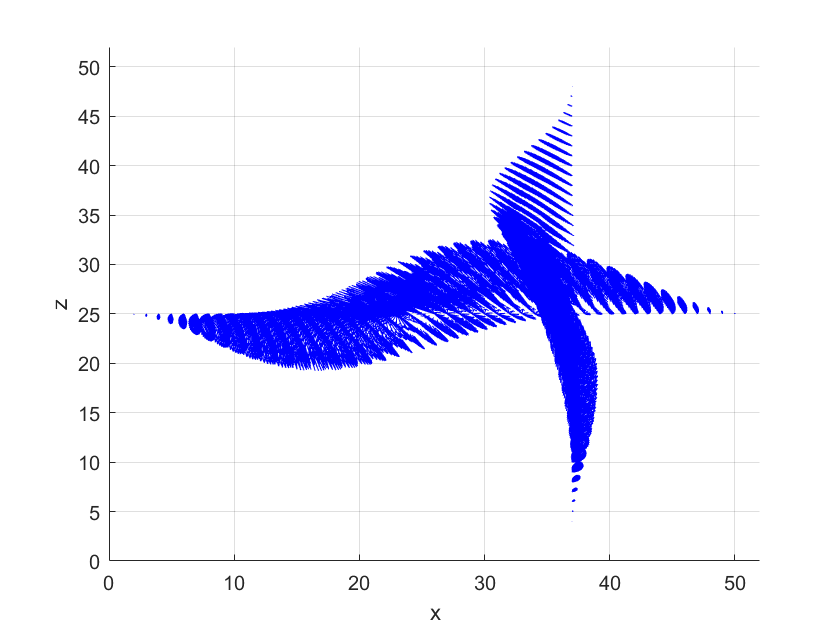}}
	\vspace{-0.5cm}
\end{figure}

\begin{figure}[H]
\vspace{-0.5cm}
\centering
	\subfigure[$\hat{\pmb{\phi}}=\pmb{id}+\hat{\pmb{u}}$ vs $\pmb{id}$ ]{\includegraphics[width=2.95cm,height=2.95cm]{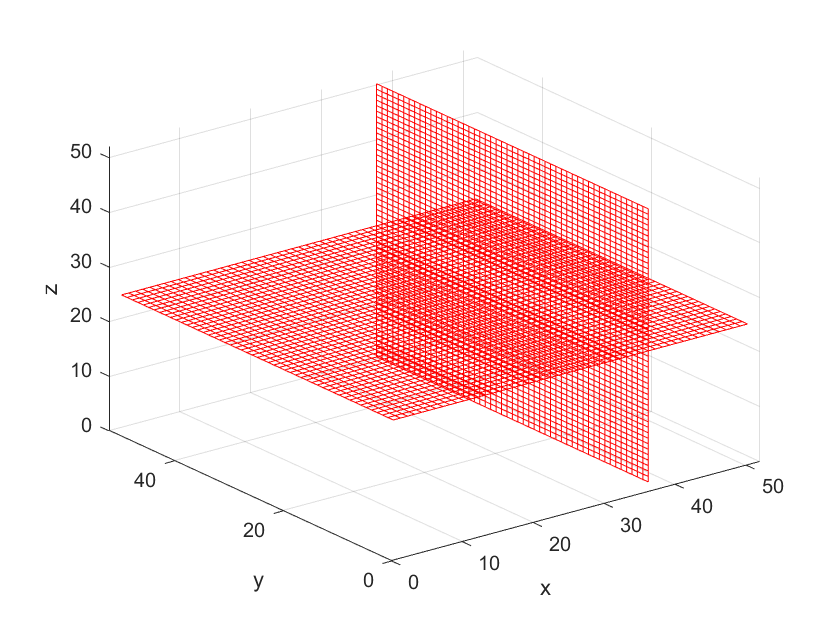}}
	\subfigure[$\hat{\pmb{\phi}}$ on xz-plain ]{\includegraphics[width=2.95cm,height=2.95cm]{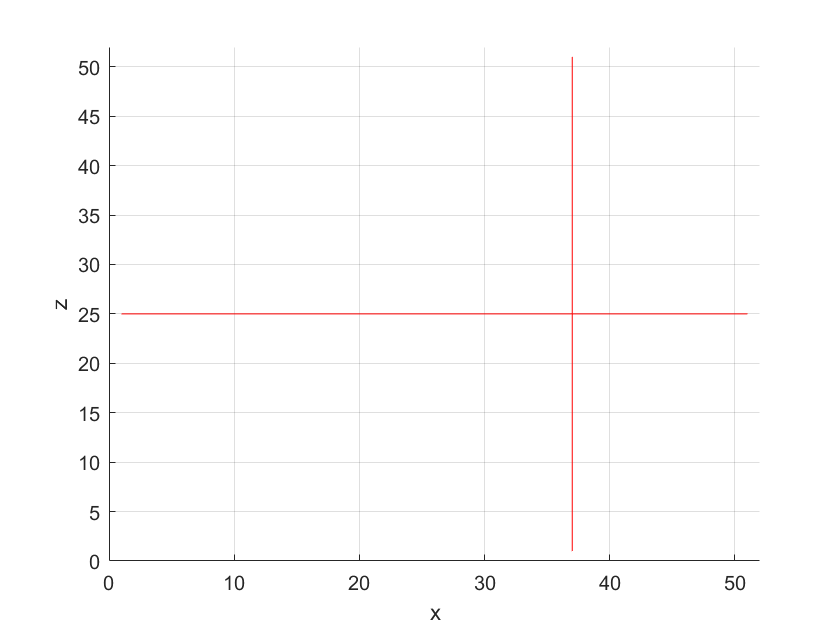}}
	\subfigure[$\hat{\pmb{u}}$]{\includegraphics[width=2.95cm,height=2.95cm]{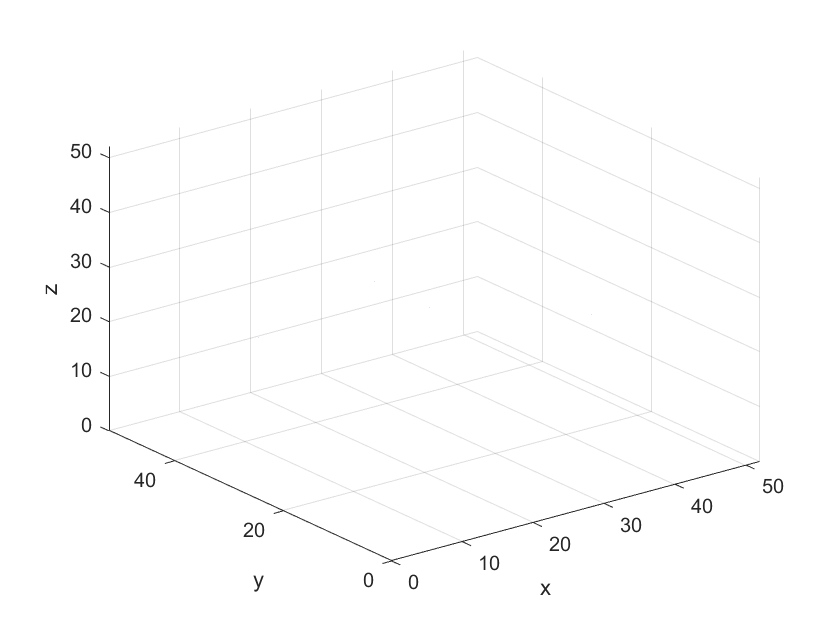}}
	\subfigure[$\hat{\pmb{u}}$ on xz-plain]{\includegraphics[width=2.95cm,height=2.95cm]{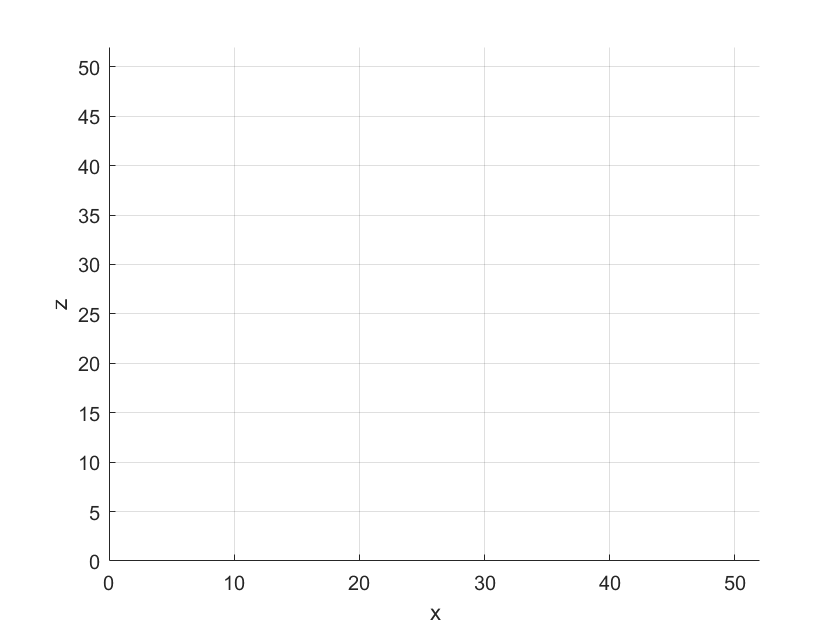}}
	\caption{Shrinking JD and Curl of $\hat{\pmb{u}}$ --- a noised $\pmb{u}$ in 3D}\label{dewave2}
	\vspace{-0.5cm}
\end{figure}

Similarly to the 2D scenario, in Figure~\ref{dewave2}, a noised example is presented. The same conclusion can be drawn, which is that the Algorithm~\ref{alg} performs as expected.
\begin{figure}[H]
	\vspace{-0.5cm}
	\centering
	\subfigure[$\hat{\pmb{\phi}}_{o}$]{\includegraphics[width=2.95cm,height=2.95cm]{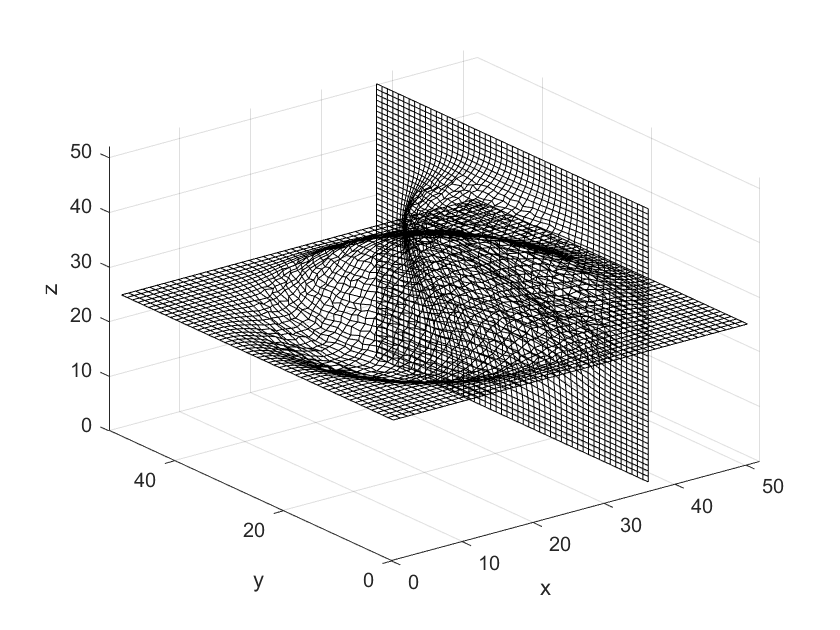}}
	\subfigure[$\hat{\pmb{\phi}}_{o}$ on xz-plain]{\includegraphics[width=2.95cm,height=2.95cm]{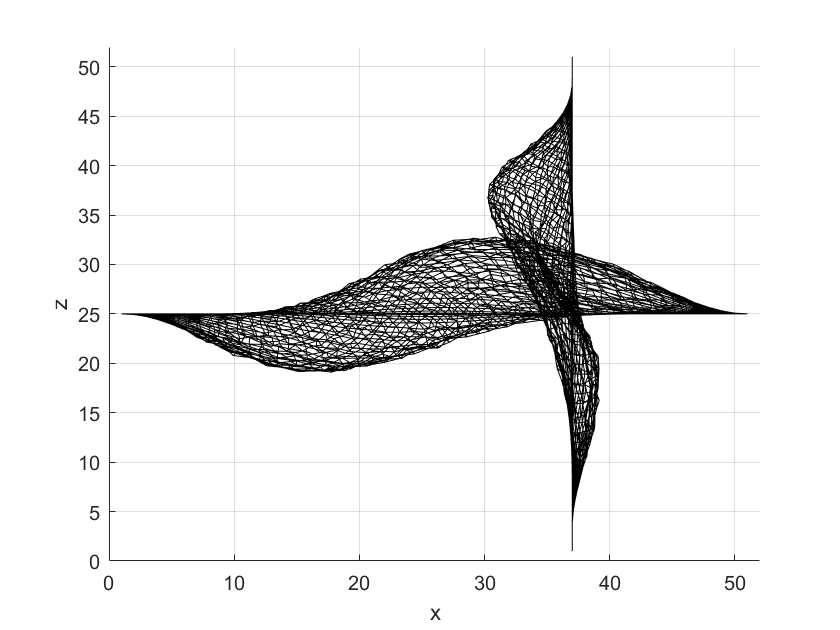}}
	\subfigure[$\hat{\pmb{u}}_{o}$]{\includegraphics[width=2.95cm,height=2.95cm]{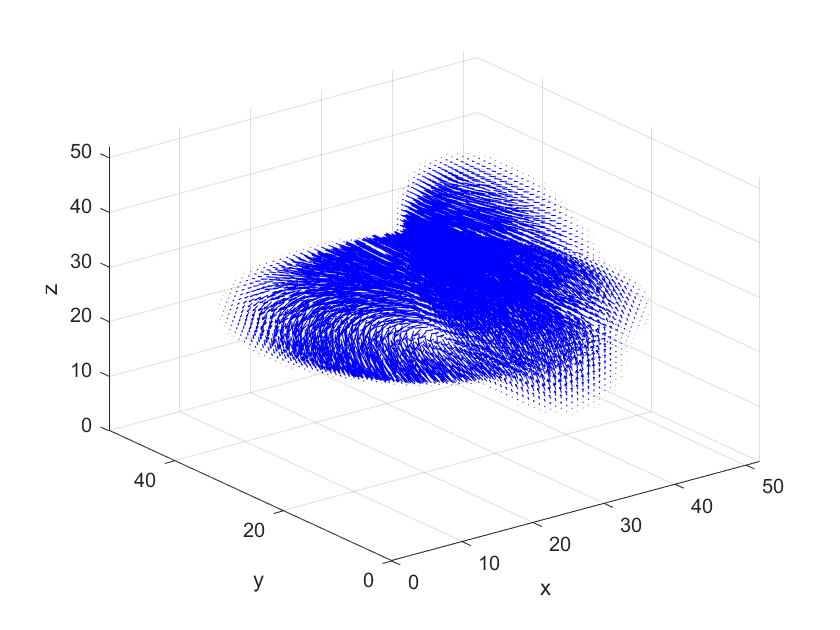}}
	\subfigure[$\hat{\pmb{u}}_{o}$ on xz-plain]{\includegraphics[width=2.95cm,height=2.95cm]{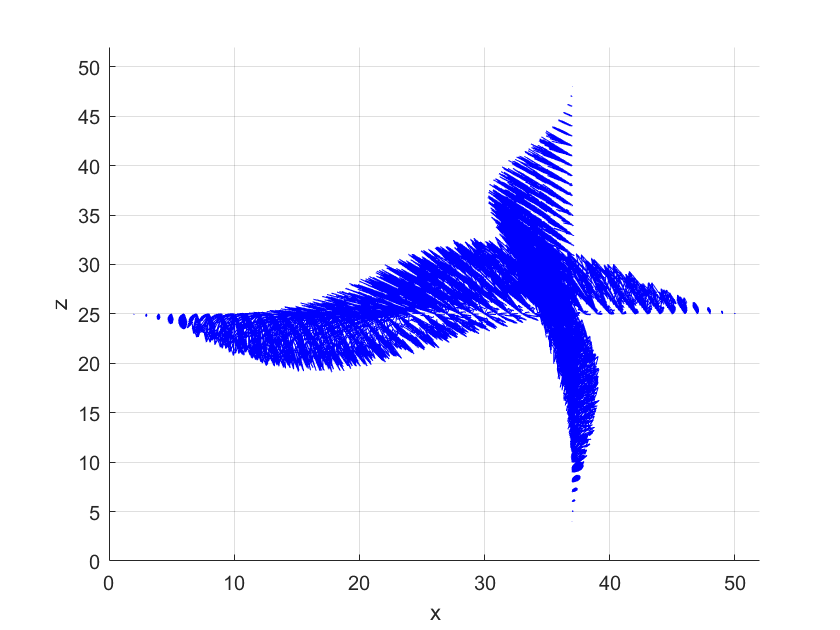}}
		\vspace{-0.5cm}
\end{figure}

\begin{table}[h]
	\begin{center}
		\begin{tabular}{|c|c|c|c|c|}
			\hline	
			\multirow{2}{*}{Solution} & \multirow{2}{*}{$ratio$} & \multicolumn{3}{c|}{max value in} \\  \cline{3-5}
			&    &$|$det$\nabla\pmb{u}^{*}|$ & $||\nabla\times\pmb{u}^{*}||_{2}$ &  $|| \pmb{u}^{*}||_{2}$ \\ \hline
			{\tt$\pmb{u}^{*}=\pmb{u}$} & $9.8399*10^{-13}$& $1.0154*10^{-6}$ &$2.2244*10^{-6}$ &  $3.0024*10^{-6}$ \\
			{\tt $\pmb{u}^{*}=\hat{\pmb{u}}$} & $9.9999*10^{-13}$& $2.8792*10^{-6}$ &$1.8388*10^{-6}$ &  $9.0450*10^{-6}$ \\
			\hline
		\end{tabular}
		\caption{Performances of Figure~\ref{dewave} and Figure~\ref{dewave2}}\label{tb3}
	\end{center}
\end{table}



Similar to the tests done in 2D, Table~\ref{tb4} shows a statistical records of 89 repeats of the similar experiment done in Figure~\ref{dewave2}. Due to the much higher computational costs in 3D, in this example, the $ratio$-tolerance is raised to $10^{-8}$, then the resulting sizes of $\left\|\pmb{u}\right\|_{L^{2}}$ is about $O(10^{-4})$. Again, as expected, Algorithm~\ref{alg} achieves mean size of $\left\|\pmb{u}\right\|_{L^{2}}\approx O(10^{-4})$ with variance about the size $O(10^{-12})$. The data forms a spike like normal distribution as well. 

\begin{table}[h]
	\begin{center}
		\begin{tabular}{|c|c|c|c|}
			\hline	
 			\multirow{2}{*}{Total 89 tests} &\multicolumn{3}{|c|}{ max values of } \\  \cline{2-4}
			 &  $|$det$\nabla\pmb{u}|$ & $||\nabla\times\pmb{u}||_{2}$ &  $|| \pmb{u}||_{2}$ \\ \hline
			 mean & $1.2677*10^{-4}$ &$2.4049*10^{-4}$ &  $5.9227*10^{-4}$ \\
			 variance & $1.2728*10^{-12}$ &$2.4049*10^{-12}$ &  $5.9227*10^{-12}$ \\
			\hline
		\end{tabular}
		\caption{Performance of 89 noised tests similar to Figure~\ref{dewave2}}\label{tb4}
	\end{center}
\end{table}

%

All the numerical tests done in this section demonstrate that minimizing $L(\pmb{u})$ based on Remark~\ref{rmk} with Algorithm~\ref{alg} is not yet able to find any counter example to the conjecture. Of course, it is known that, regardless the number of tests in checking counter example, the full generality is reachable. On one hand, Remark~\ref{rmk} presents an idea about how to find a counter example to the conjecture; on the other hand, it is also essential to analytically check whether the conjecture is true or false. This leads to the next section.

\section{An Intermediate Step to the Uniqueness Conjecture}
In this section, the analysis includes (1) referred as the simple case, it is argued that if a diffeomorphism $\pmb{\phi}$ is close to the $\pmb{id}$ in ${H_{0}^{2}(\mathrm{\Omega})}$ within certain discrepancy, then $\pmb{\phi}$ is $\pmb{id}$; (2) based on the simple case, an intermediate step to the conjecture is also provided and referred as the semi-general case. 
\begin{theorem}\label{simple}
	Let $\pmb{\phi}(\pmb{\omega})=\pmb{id}+\pmb{u}=<x+u_1(x,y,z),y+u_2(x,y,z),z+u_3(x,y,z)>$ and $\pmb{\psi}(\pmb{\omega})=\pmb{id}=<x,y,z>$ in ${H_{0}^{2}(\mathrm{\Omega})}$ such that $\left\|\pmb{u}\right\|_{H_{0}^{2}(\mathrm{\Omega})} < \epsilon$ holds for the choice of $\epsilon\in(0,\min\{1, 1/\sqrt{C}\})\subset \mathbb{R}$ where $C$ is determined by the $Poincare's$ inequality. If $\pmb{\phi}$ and $\pmb{\psi}$ satisfy (\ref{UniqueCon1}) and (\ref{UniqueCon2}), then $\pmb{\phi}=\pmb{\psi}$.
\end{theorem}
 \begin{proof} From $\left\|\pmb{u}\right\|_{H_{0}^{2}(\mathrm{\Omega})} < \epsilon$, it is known
\begin{equation}\label{sobolev}
	\left\{
	\begin{aligned}
		\left\|\pmb{u}\right\|_{L^{2}}& <\epsilon,\\
		\left\|\nabla \pmb{u}\right\|_{L^{2}}&< \epsilon,\\
		\left\|\mathrm{\Delta}\pmb{u}\right\|_{L^{2}}&< \epsilon.\\
	\end{aligned}\right.
\end{equation} 
From (\ref{Jd_phi}), it is denoted
\begin{equation*}
	\text{det}\nabla\pmb{\phi}=1+\nabla\cdot\pmb{u}-\mathcal{F}(\pmb{u}), \text{ where }\mathcal{F}(\pmb{u})=-[\text{det}\nabla\pmb{u}+\text{Tail}(\pmb{
		u})].
\end{equation*}
According to (\ref{UniqueCon1}) and (\ref{UniqueCon2}), it leads to
\begin{equation*}
	\left\{
	\begin{aligned}
	&0=\text{det}\nabla\pmb{\phi}-\text{det}\nabla\pmb{\psi}= 1+\nabla\cdot\pmb{u}- \mathcal{F}(\pmb{u})-1=\nabla\cdot\pmb{u}-\mathcal{F}(\pmb{u}),\\ 
	&\pmb{0}=\nabla\times\pmb{\phi}-\nabla\times\pmb{\psi}=
	\begin{pmatrix}
		u_{3y}-u_{2z}\\ 
		u_{1z}-u_{3x}\\ 
		u_{1y}-u_{2x} \notag 
	\end{pmatrix}-\pmb{0}= \nabla\times\pmb{u}.
	\end{aligned}\right.
\end{equation*}
\begin{equation*}
	\Rightarrow\left\{
	\begin{aligned}
		&\nabla\cdot\pmb{u} =\mathcal{F}(\pmb{u}),\\
		&\nabla\times\pmb{u} = \pmb{0}.
	\end{aligned}\right.
\end{equation*}
\begin{equation*}
	\Rightarrow\left\{
	\begin{aligned}
		\mathrm{\Delta} u_1& =(\nabla\cdot\pmb{u})_{x}-(\nabla\times\pmb{u})_{3y}+(\nabla\times\pmb{u})_{2z}=\mathcal{F}(\pmb{u})_{x},\\
		\mathrm{\Delta} u_2& =(\nabla\cdot\pmb{u})_{y}+(\nabla\times\pmb{u})_{3x}-(\nabla\times\pmb{u})_{1z}=\mathcal{F}(\pmb{u})_{y},\\
		\mathrm{\Delta} u_3& =(\nabla\cdot\pmb{u})_{z}-(\nabla\times\pmb{u})_{2x}+(\nabla\times\pmb{u})_{1y}=\mathcal{F}(\pmb{u})_{z},\\
	\end{aligned}\right.
\end{equation*}
\begin{equation*}
	\Rightarrow \mathrm{\Delta} \pmb{u} = 	<\mathcal{F}(\pmb{u})_{x},\mathcal{F}(\pmb{u})_{y},\mathcal{F}(\pmb{u})_{z}>= {\nabla_{\pmb{\omega}}\mathcal{F}(\pmb{u})}.
\end{equation*}
The dominating terms of ${\mathcal{F}(\pmb{u})}$ at (\ref{Jd_phi}) are products of only the first partial derivatives of $\pmb{u}$, so the dominating terms of $\nabla_{\pmb{\omega}}{\mathcal{F}(\pmb{u})}$ in
\begin{equation}\label{laplace}
	\left\|\mathrm{\Delta}\pmb{u} \right\|_{L^{2}}=(\int_{\mathrm{\Omega}}|\mathrm{\Delta} \pmb{u}|^2)^{\frac{1}{2}}=(\int_{\mathrm{\Omega}}| \nabla_{\pmb{\omega}}{\mathcal{F}(\pmb{u})}|^2)^{\frac{1}{2}}=\left\|\nabla_{\pmb{\omega}}{\mathcal{F}(\pmb{u})} \right\|_{L^{2}}
\end{equation}
are the terms in products of the first and second partial derivatives of $\pmb{u}$. By (\ref{sobolev}) and (\ref{laplace}), it forms
\begin{equation}\label{lap_improve}
	\left\|\mathrm{\Delta}\pmb{u} \right\|_{L^{2}}=\left\|\nabla_{\pmb{\omega}}{\mathcal{F}(\pmb{u})} \right\|_{L^{2}}\le\left\|\pmb{u}\right\|_{L^{2}}\left\|\mathrm{\Delta}\pmb{u} \right\|_{L^{2}}<\epsilon\cdot\epsilon=\epsilon^{2}.
\end{equation}
Next, an inequality for $\left\|\pmb{u}\right\|_{L^{2}}$ is established. Since $\pmb{u}=0\text{ on } \partial\mathrm{\Omega}$, by $Green's$ formula, it can be derived to
\begin{equation}\label{eq217}
	\int_{\mathrm{\Omega}}|\nabla \pmb{u}|^2=|\int_{\mathrm{\Omega}}\pmb{u} \cdot \mathrm{\Delta}\pmb{u}|.
\end{equation}
Applying the $Cauchy-Schwarz$ inequality and properties of integration to the right-hand-side of (\ref{eq217}) to get
\begin{equation*}
	\int_{\mathrm{\Omega}}|\nabla \pmb{u}|^2=|\int_{\mathrm{\Omega}}\pmb{u} \cdot \mathrm{\Delta}\pmb{u}|\leq\int_{\mathrm{\Omega}}|\pmb{u} ||\mathrm{\Delta}\pmb{u}|\leq\left\|\pmb{u}\right\|_{L^{2}}\left\|\mathrm{\Delta}\pmb{u}\right\|_{L^{2}}
\end{equation*}
\begin{equation}\label{RHS}
	\Rightarrow \left\|\nabla \pmb{u}\right\|_{L^{2}}^{2}\leq\left\|\pmb{u}\right\|_{L^{2}}\left\|\mathrm{\Delta}\pmb{u}\right\|_{L^{2}}.
\end{equation}
Applying the $Poincare$'s inequality to the left-hand-side of (\ref{eq217}), $\exists$ $0<C\in\mathbb{R}$, such that
\begin{equation}\label{eq219}
	\left\|\pmb{u}\right\|_{L^{2}}^{2}\leq C \left\|\nabla \pmb{u}\right\|_{L^{2}}^{2}= C \int_{\mathrm{\Omega}}|\nabla \pmb{u}|^2.
\end{equation}
Then, combining (\ref{RHS}) and (\ref{eq219}) to have
\begin{equation}\label{u_improve}
	\left\|\pmb{u}\right\|_{L^{2}}^{2}\leq C \left\|\nabla \pmb{u}\right\|_{L^{2}}^{2}\leq C \left\|\pmb{u}\right\|_{L^{2}}\left\|\mathrm{\Delta}\pmb{u}\right\|_{L^{2}}
\end{equation}
\begin{equation}\label{LHS}
	\Rightarrow \left\|\pmb{u}\right\|_{L^{2}}\leq C \left\|\mathrm{\Delta}\pmb{u}\right\|_{L^{2}}< C \epsilon^2.
\end{equation}
As for $\left\|\nabla \pmb{u}\right\|_{L^{2}}$, by (\ref{lap_improve}), (\ref{RHS}) and (\ref{LHS}), it can be bounded as 
\begin{equation}\label{grad_improve}
	\left\|\nabla \pmb{u}\right\|_{L^{2}}\leq(\left\|\pmb{u}\right\|_{L^{2}}\left\|\mathrm{\Delta}\pmb{u}\right\|_{L^{2}})^{\frac{1}{2}}
\end{equation}
\begin{equation*}
	\Rightarrow  \left\|\nabla \pmb{u}\right\|_{L^{2}}<(C \epsilon^2 \cdot\epsilon^{2})^{\frac{1}{2}}=C^{\frac{1}{2}}\epsilon^{2}.
\end{equation*}
So, $\pmb{u}$ from (\ref{sobolev}) also satisfies 
\begin{equation}\label{u_improve_0}
	\left\{
	\begin{aligned}
		\left\|\pmb{u}\right\|_{L^{2}}& <C\epsilon^2,\\
		\left\|\nabla \pmb{u}\right\|_{L^{2}}& < C^{\frac{1}{2}}\epsilon^2,\\
		\left\|\mathrm{\Delta}\pmb{u}\right\|_{L^{2}}& < \epsilon^2.\\
	\end{aligned}\right.
\end{equation}
Now, treating the procedure of (\ref{sobolev}) to (\ref{u_improve_0}) as the step $n=0$, and repeat it again with replacing (\ref{sobolev}) by (\ref{u_improve_0}) to see how the step $n=1$ looks. So, plugging (\ref{u_improve_0}) into the first derivative of $\pmb{u}$ to (\ref{lap_improve}), (\ref{u_improve}) and (\ref{grad_improve}), respectively, it forwards to
\begin{equation}\label{u_improve_01}
	\left\|\mathrm{\Delta}\pmb{u}\right\|_{L^{2}} =\left\|\nabla_{\pmb{\omega}}{\mathcal{F}(\pmb{u})} \right\|_{L^{2}}<\epsilon\cdot C^{\frac{1}{2}}\epsilon^2=C^{\frac{1}{2}}\epsilon^3,
\end{equation}
\begin{equation}\label{u_improve_02}
	\left\|\pmb{u}\right\|_{L^{2}}\leq C \left\|\mathrm{\Delta}\pmb{u}\right\|_{L^{2}}< C \cdot C^{\frac{1}{2}} \epsilon^3=C^{(1+\frac{1}{2})}\epsilon^3
\end{equation}
and 
\begin{equation}\label{u_improve_03}
	\left\|\nabla \pmb{u}\right\|_{L^{2}}\leq(\left\|\pmb{u}\right\|_{L^{2}}\left\|\mathrm{\Delta}\pmb{u}\right\|_{L^{2}})^{\frac{1}{2}}<(C^{(1+\frac{1}{2})} \epsilon^3 \cdot C^{\frac{1}{2}}\epsilon^{3})^{\frac{1}{2}}=C\epsilon^{3}, 
\end{equation}
respectively. Combining (\ref{u_improve_03}), (\ref{u_improve_01}) and (\ref{u_improve_02}) into the $n=1$ case,
\begin{equation}\label{u_improve_1}
	\left\{
	\begin{aligned}
		\left\|\pmb{u}\right\|_{L^{2}}& <C^{1+\frac{1}{2}*(n=1)}\epsilon^{2+1*(n=1)},\\
		\left\|\nabla \pmb{u}\right\|_{L^{2}}& < C^{\frac{1}{2}+\frac{1}{2}*(n=1)}\epsilon^{2+1*(n=1)},\\
		\left\|\mathrm{\Delta}\pmb{u}\right\|_{L^{2}}& < C^{(0+\frac{1}{2}*(n=1))}\epsilon^{2+1*(n=1)}.\\
	\end{aligned}\right.
\end{equation}
Please note that from (\ref{u_improve_01}) to (\ref{u_improve_1}), with one power raised on $\epsilon$, the upper bounds are compensated with a half-power raised on $C$ from the $Poincare$'s inequality. Now, taking the format of (\ref{u_improve_1}), suppose that it is true at $n=k$ case, i.e., 
\begin{equation}\label{u_improve_k}
	\left\{
	\begin{aligned}
		\left\|\pmb{u}\right\|_{L^{2}}& <C^{(1+\frac{k}{2})}\epsilon^{(2+k)},\\
		\left\|\nabla \pmb{u}\right\|_{L^{2}}& < C^{(\frac{1}{2}+\frac{k}{2})}\epsilon^{(2+k)},\\
		\left\|\mathrm{\Delta}\pmb{u} \right\|_{L^{2}}& <C^{(0+\frac{k}{2})}\epsilon^{(2+k)}.\\
	\end{aligned}\right.
\end{equation}
We want to see an inductive step forward at $n=k+1$ is true or not. Affirmatively, this can be achieved by plugging (\ref{u_improve_k}) to (\ref{lap_improve}), (\ref{u_improve}) and (\ref{grad_improve}), respectively, then, it can be carried to
\begin{equation}\label{lap_improve_k1}
	\left\|\mathrm{\Delta}\pmb{u} \right\|_{L^{2}}=\left\|\nabla_{\pmb{\omega}}{\mathcal{F}(\pmb{u})} \right\|_{L^{2}}<\epsilon \cdot C^{(\frac{1}{2}+\frac{k}{2})}\epsilon^{(2+k)}=C^{(0+\frac{k+1}{2})}\epsilon^{(2+k+1)},
\end{equation}
\begin{equation}\label{u_improve_k1}
	\left\|\pmb{u}\right\|_{L^{2}}\leq C \left\|\mathrm{\Delta}\pmb{u}\right\|_{L^{2}}< C \cdot C^{(0+\frac{k+1}{2})}\epsilon^{(2+k+1)}=C^{(1+\frac{k+1}{2})}\epsilon^{(2+k+1)},
\end{equation}
and
\begin{equation}\label{grad_improve_k1}
	\left\|\nabla \pmb{u}\right\|_{L^{2}}<(C^{(1+\frac{k+1}{2})} \epsilon^{(2+k+1)} \cdot C^{(0+\frac{k+1}{2})}\epsilon^{(2+k+1)})^{\frac{1}{2}}=C^{(\frac{1}{2}+\frac{k+1}{2})}\epsilon^{(2+k+1)},
\end{equation}
respectively. Similarly, combining (\ref{u_improve_k1}), (\ref{grad_improve_k1}) and (\ref{lap_improve_k1}) into the following, 
\begin{equation}\label{u_improve_1k+1}
	\left\{
	\begin{aligned}
		\left\|\pmb{u}\right\|_{L^{2}}& <C^{(1+\frac{k+1}{2})}\epsilon^{(2+k+1)},\\
		\left\|\nabla \pmb{u}\right\|_{L^{2}}& < C^{(\frac{1}{2}+\frac{k+1}{2})}\epsilon^{(2+k+1)},\\
		\left\|\mathrm{\Delta}\pmb{u} \right\|_{L^{2}}& <C^{(0+\frac{k+1}{2})}\epsilon^{(2+k+1)},\\
	\end{aligned}\right.
\end{equation}
which is the immediate inductive step for $n=k+1$. Therefore, the system of inequalities $(\ref{u_improve_1k+1})$ can be is iterated to reduce the bound of $\left\|\pmb{u}\right\|_{L^{2}}$ for every increment of $k=k+1$ up to any $n\in \mathbb{N}$. That is 
\begin{equation}\label{u_improve_n}
	\left\{
	\begin{aligned}
		\left\|\pmb{u}\right\|_{L^{2}}& <C^{(1+\frac{n}{2})}\epsilon^{(2+n)},\\
		\left\|\nabla \pmb{u}\right\|_{L^{2}}& < C^{(\frac{1}{2}+\frac{n}{2})}\epsilon^{(2+n)},\\
		\left\|\mathrm{\Delta}\pmb{u} \right\|_{L^{2}}& <C^{(0+\frac{n}{2})}\epsilon^{(2+n)}.\\
	\end{aligned}\right.
\end{equation}
Hence, for $\pmb{u}$ in (\ref{sobolev}) is also satisfying $\left\|\pmb{u}\right\|_{L^{2}} <C^{(1+\frac{n}{2})}\epsilon^{(2+n)}$ in (\ref{u_improve_n}), where $C^{(1+\frac{n}{2})}\epsilon^{(2+n)}$ converges to $0$ by the choice of $0<\epsilon<\min\{1, 1/\sqrt{C}\}$ as $ n \longrightarrow \infty$. So, $\left\|\pmb{\phi}-\pmb{\psi} \right\|_{L^{2}}=\left\|\pmb{u} \right\|_{L^{2}}\longrightarrow 0$ as $ n \longrightarrow \infty$, which implies 
$\pmb{\phi} \equiv \pmb{\psi}=\pmb{id}$ on $\mathrm{\Omega}$.
\end{proof}

The analysis above describes an approach to the uniqueness problem based on the simple case which one smooth transformation is close $\pmb{id}$ and the other one is $\pmb{id}$. The general uniqueness problem defined by (\ref{UniqueCon1}) to (\ref{UniqueCon2}) remains open. But, an interesting intermediate step is to show that, for any two sufficiently close $\pmb{\phi}$ and $\pmb{\psi}$, a similar argument can be applied. Section \ref{VP} indicates that a solution by VP should land in a Lie group of ${H_{0}^{2}(\mathrm{\Omega})}$ whose group operation is the function composition. To formulate the analysis in this context, we relax the generality a bit by adopting the definition of small deformation in \cite{Joshi}, i.e., for $\pmb{T}=\pmb{id}+\pmb{u}\in H^{2}_{0}(\mathrm{\Omega})$, whose function composition is defined by
\begin{equation}\label{small}
	\pmb{T}_2 \circ \pmb{T}_1=\pmb{T}_{1}+\pmb{u}_{2}=\pmb{id}+\pmb{u}_{1}+\pmb{u}_{2}=\pmb{id}+\pmb{u}_{2}+\pmb{u}_{1}=\pmb{T}_{2}+\pmb{u}_{1}=\pmb{T}_1 \circ \pmb{T}_2.
\end{equation}
\begin{lemma}\label{lemma1}
	For $\pmb{T}=\pmb{id}+\pmb{u}\in H^{2}_{0}(\mathrm{\Omega})$ satisfies (\ref{small}), denote $\pmb{T}^{-1}=\pmb{id}+\pmb{v}$, then $\pmb{T}^{-1}=\pmb{id}-\pmb{u}$.
\end{lemma}
\begin{proof}
	\begin{equation*}
			\pmb{id}=\pmb{T}^{-1}\circ\pmb{T}	=\pmb{id}+\pmb{u}+\pmb{v}\Rightarrow\pmb{v}=-\pmb{u}\Rightarrow\pmb{T}^{-1}=\pmb{id}+\pmb{v}=\pmb{id}-\pmb{u}.
	\end{equation*}
\end{proof}
\begin{lemma}\label{lemma2}
	For $\pmb{T}=\pmb{id}+\pmb{u}\in H^{2}_{0}(\mathrm{\Omega})$ satisfies (\ref{small}), denote $\pmb{T}^{-1}=\pmb{id}+\pmb{v}$, then $\nabla\times\pmb{T}^{-1}=-\nabla\times\pmb{T}$.
\end{lemma}
\begin{proof}
	By definition of curl and (\ref{small}), it straightforwards to 
	\begin{equation*}
		\begin{aligned}
			0&=\nabla\times\pmb{id}=\nabla\times(\pmb{T}^{-1}\circ\pmb{T})\\
			&=\nabla\times(\pmb{id}+\pmb{u}+\pmb{v})=\nabla\times(\pmb{id}+\pmb{v}+\pmb{u})\\
			&=\nabla\times\pmb{T}^{-1}+\nabla\times\pmb{u}=\nabla\times\pmb{T}^{-1}+\nabla\times\pmb{T},
		\end{aligned}
	\end{equation*}
	\begin{equation*}
	\Rightarrow\nabla\times\pmb{T}^{-1}=-\nabla\times\pmb{T}.
\end{equation*}
\end{proof}

With lemma \ref{lemma1} and \ref{lemma2}, one may find the following particular result, which we refer it as the semi-general case, due to its adoption of function composition among the small deformations.   

\begin{theorem}\label{semiG}
Let $\pmb{\phi}=\pmb{id}+\pmb{u}$, $\pmb{\psi}=\pmb{id}+\pmb{v}\in H^{2}_{0}(\mathrm{\Omega})\in H_{0}^{2}(\mathrm{\Omega})$. Suppose $\pmb{\phi}$ and $\pmb{\psi}$ satisfy (\ref{UniqueCon1}), (\ref{UniqueCon2}) and (\ref{small}). $\exists$ $\epsilon>0$, such that, if $\left\|\pmb{u}-\pmb{v}\right\|_{H_{0}^{2}(\mathrm{\Omega})}<\epsilon$, then $\pmb{\phi}=\pmb{\psi}$.
\end{theorem}
\begin{proof}
From (\ref{UniqueCon1}), one may find
\begin{equation*}
		\text{det}\nabla\pmb{\phi}=\text{det}\nabla\pmb{\psi}\Rightarrow \frac{\text{det}\nabla\pmb{\phi}}{\text{det}\nabla\pmb{\psi}}=1\Rightarrow\text{det}\nabla\pmb{\psi}^{-1}\text{det}\nabla\pmb{\phi}=1,
\end{equation*}	
\begin{equation}\label{group1}
	\Rightarrow\text{det}\nabla(\pmb{\psi}^{-1}\circ\pmb{\phi})=\text{det}\nabla\pmb{id}.
\end{equation}	
Next, from (\ref{UniqueCon2}), by (\ref{lemma2}), one may also find
\begin{equation*}
	\nabla\times\pmb{\phi}=\nabla\times\pmb{\psi}\Rightarrow\nabla\times\pmb{\phi}-\nabla\times\pmb{\psi}=0\Rightarrow\nabla\times\pmb{\phi}+\nabla\times\pmb{\psi}^{-1}=0,
\end{equation*}
\begin{equation*}
	\Rightarrow\nabla\times(\pmb{id}+\pmb{u}-\pmb{v})=0,
\end{equation*}
\begin{equation}\label{group2}
	\Rightarrow\nabla\times(\pmb{\psi}^{-1}\circ\pmb{\phi})=\nabla\times\pmb{id}.
\end{equation}
By Lemma \ref{lemma1}, denote $\pmb{w}=\pmb{u}-\pmb{v}$, it leads to
\begin{equation}\label{group3}
\pmb{\psi}^{-1}\circ\pmb{\phi}=\pmb{id}+\pmb{w}=\pmb{id}+\pmb{u}-\pmb{v}\Rightarrow\left\|\pmb{w}\right\|=\left\|\pmb{u}-\pmb{v}\right\|<\epsilon.
\end{equation}
Finally, (\ref{group1}), (\ref{group2}) and (\ref{group3}) indicates that $\pmb{\psi}^{-1}\circ\pmb{\phi}$ satisfy the conditions of Theorem \ref{simple}, therefore, 
\begin{equation}\label{group4}
\pmb{\psi}^{-1}\circ\pmb{\phi}\equiv\pmb{id}
\Rightarrow \pmb{\psi}^{-1}\circ\pmb{\phi}=\pmb{\psi}^{-1}\circ\pmb{\psi} \Rightarrow \pmb{\phi}=\pmb{\psi},
\end{equation}
as desired.
\end{proof}

\section*{Conclusion}
In summary, the Conjecture is still open. This paper (1) provides an experimental strategy to numerically check for counter examples to the uniqueness conjecture; (2) this paper describes an intermediate step to argue for the uniqueness conjecture, which considers the two smooth transformations that are close to each other under the small deformation category. The general uniqueness conjecture remains open. An approach to bypass the restriction of small deformation would potentially demythologize the Conjecture. 

%
\bibliographystyle{amsplain}

\begin{thebibliography}{12}
\bibitem{Bauer}M.~Bauer, S. Joshi and K.~Modin, Diffeomorphic density matching by optimal information transport, SIAM Journal on Imaging Sciences, \textbf{8}{3} 1718–1751 (2015)	

\bibitem{Cai} X.~Cai, D.~Fleitas, B.~Jiang and G.~Liao, Adaptive grid generation based on the least-squares finite elements method, Computers and Mathematics with Applications, \textbf{48}, 1007-1085 (2004)	 

\bibitem{Castillo}J.~Castillo, S.~Steinberg, and P.~Roach, Parameter estimation in variational grid generation, Appl. Math. Comput., \textbf{155}, 155-177 (1988)	

\bibitem {ChenY}Y.~Chen and X.~Ye, Inverse Consistent Deformable Image Registraition. Springer Verlag, 419-440 (2010).

\bibitem{ChenXi}X.~Chen and G.~Liao, New Variational Method of Grid Generation with Prescribed Jacobian determinant and Prescribed Curl, arxiv.org/pdf/1507.03715 (2015)

\bibitem {XiChen3}
X.~Chen and G.~Liao,
New method of averaging diffeomorphisms based on Jacobian determinant and curl vector, arxiv.org/abs/1611.03946  (2016)

\bibitem{DacMos} B.~Dacoragna and J.~Moser, On A Partial Differential Equation Involving the Jacobian determinant, Ann.Inst H Poincaré, \textbf{7}, 1-26 (1990)	

\bibitem{Gu}X.~Gu and S.T.~Yau, Computational Conformal Geometry, International Press of Boston, Inc (2008)


\bibitem{Grajewski}
M.~Grajewski, M.~Koster and S.~Turek, Mathematical and Numerical Analysis of a Robust and Efficient Grid Deformation Method in the Finite Element Context, \emph{SIAM Journal on Scientific Computing},  \textbf{31}{2}, 1539-1557 (2009)

\bibitem{Huang} W.~Huang and W.~Sun, Variational mesh adaptation II: Error estimates and monitor functions, \emph{J. Comput. Phys}, \textbf{184}, 619-648 (2003)


\bibitem{Joshi} S.~Joshi, B.~Davis, M.~Jomier and G.~Gerig, Unbiased Diffeomorphic Atlas Construction for Computational Anatomy, NeuroImage, \textbf{23}, 151-160 (2004)


\bibitem{Liao} G.~Liao, X.~Cai, J.~Liu, X.~Luo, J.~Wang, J.~Xue, Construction of differentiable transformations, Applied Math.Letters,\textbf{22}, 543-1548 (2009)	

\bibitem {Liseikin}V.~Liseikin, Grid Generation Method, Springer Press. (1999)

\bibitem{Moser} J.~Moser, Volume elements of a Riemann manifold, Trans. AMS, \textbf{120}, 155-177 (1965).	

\bibitem{Zhou2} Z.~Zhou and G.~Liao, Construction of Diffeomorphisms with Prescribed Jacobian Determinant and Curl, International Conference on Geometry and Graphics, Proceedings (in press) (2022)

\bibitem{Zhou} Z.~Zhou, Image Analysis Based on Differential Operators with Applications to Brain MRIs, \emph{Ph.D. Dissertation}, University of Texas at Arlington (2019)
	
\end{thebibliography}

\newpage
\section{Appendix}\label{app1}
\subsection*{Gradient of $L(\pmb{u})$ in (\ref{Lu}) with respect to control function $\pmb{F}$}
A minimizer to $L(\pmb{u})$ (\ref{Lu}) is necessarily having vanishing variational gradients of $L_{1}(\pmb{u})$ and $L_{2}(\pmb{u})$ with respect to the control function $\pmb{F}$ as in (\ref{ssdconst}). Next, denote $P=\nabla\cdot\pmb{u}+\text{det}\nabla\pmb{u}+\text{Tail}(\pmb{u})$ and $\pmb{Q}=\nabla \times\pmb{u}$, then, for all $\delta\pmb{F}$ vanishing on $\partial\mathrm{\Omega}$, (\ref{Lu}) leads to

\begin{equation*}
	\begin{aligned}
		\delta L(\pmb{u})&= \delta L_{1}(\pmb{u})+\delta L_{2}(\pmb{u})=\frac{1}{2}\int_{\mathrm{\Omega}} \delta[(\nabla\cdot\pmb{u}+\text{det}\nabla\pmb{u}+\text{Tail}(\pmb{u}))^2+\delta|\nabla \times\pmb{u}|^2] d\pmb{\omega},\\
	\end{aligned}
\end{equation*}		
\begin{equation*}
	\begin{aligned}
	\Rightarrow\delta L_{1}&(\pmb{u})=\int_{\mathrm{\Omega}} [(\nabla\cdot\pmb{u}+\text{det}\nabla\pmb{u}+\text{Tail}(\pmb{u}))(\delta\nabla\cdot\pmb{u}+\delta\text{det}\nabla\pmb{u}+\delta\text{Tail}(\pmb{u}))] d\pmb{\omega}\\
	=&\int_{\mathrm{\Omega}} [P\nabla\cdot\delta\pmb{u}+P\delta\text{det}\nabla\pmb{u}+P\delta\text{Tail}(\pmb{u}))] d\pmb{\omega}\\
	=&\int_{\mathrm{\Omega}} [P\nabla\cdot\delta\pmb{u}+P(\delta u_{1x}u_{2y}u_{3z}+u_{1x}\delta u_{2y}u_{3z}+u_{1x}u_{2y}\delta u_{3z}\\
	&\quad-\delta u_{1x}u_{2z}u_{3y}-u_{1x}\delta u_{2z}u_{3y}-u_{1x}u_{2z}\delta u_{3y}\\
	&\quad-\delta u_{1y}u_{3z}u_{2x}-u_{1y}\delta u_{3z}u_{2x}-u_{1y}u_{3z}\delta u_{2x}\\
	&\quad+\delta u_{1y}u_{3x}u_{2z}+u_{1y}\delta u_{3x}u_{2z}+u_{1y}u_{3x}\delta u_{2z}\\
	&\quad+\delta u_{1z}u_{2x}u_{3y}+u_{1z}\delta u_{2x}u_{3y}+u_{1z}u_{2x}\delta u_{3y}\\
	&\quad-\delta u_{1z}u_{2y}u_{3x}-u_{1z}\delta u_{2y}u_{3x}-u_{1z}u_{2y}\delta u_{3x})+P\delta\text{Tail}(\pmb{u})]\\
	=&\int_{\mathrm{\Omega}} [P\nabla\cdot\delta\pmb{u}+P\begin{pmatrix}
		u_{ 2y}u_{ 3z}-u_{ 3y}u_{ 2z} \\ 
		u_{ 3x}u_{ 2z}-u_{ 2x}u_{ 3z} \\ 
		u_{ 2x}u_{ 3y}-u_{ 2y}u_{ 3x} \notag   
	\end{pmatrix}\cdot  \nabla \delta u_1\\
	+P&\begin{pmatrix}
		u_{ 3y}u_{ 1z}-u_{ 1y}u_{ 3z} \\ 
		u_{ 1x}u_{ 3z}-u_{ 1z}u_{ 3x} \\ 
		u_{ 3x}u_{ 1y}-u_{ 1x}u_{ 3y} \notag   
	\end{pmatrix}\cdot  \nabla \delta u_2
	+P\begin{pmatrix}
		u_{ 1y}u_{ 2z}-u_{ 2y}u_{ 1z} \\ 
		u_{ 2x}u_{ 1z}-u_{ 1x}u_{ 2z} \\ 
		u_{ 1x}u_{ 2y}-u_{ 2x}u_{ 1y} \notag   
	\end{pmatrix}\cdot\nabla \delta u_3\\
	+P&\delta(u_{1x}u_{2y}+ u_{1x}u_{3z}+ u_{2y}u_{3z}
	- u_{1y}u_{2x}- u_{1z}u_{3x}- u_{2z}u_{3y})]d\pmb{\omega}\\
	=&\int_{\mathrm{\Omega}} [P\nabla\cdot\delta\pmb{u}+P\begin{pmatrix}
		u_{ 2y}u_{ 3z}-u_{ 3y}u_{ 2z} \\ 
		u_{ 3x}u_{ 2z}-u_{ 2x}u_{ 3z} \\ 
		u_{ 2x}u_{ 3y}-u_{ 2y}u_{ 3x} \notag   
	\end{pmatrix}\cdot  \nabla \delta u_1\\
	+P&\begin{pmatrix}
		u_{ 3y}u_{ 1z}-u_{ 1y}u_{ 3z} \\ 
		u_{ 1x}u_{ 3z}-u_{ 1z}u_{ 3x} \\ 
		u_{ 3x}u_{ 1y}-u_{ 1x}u_{ 3y} \notag   
	\end{pmatrix}\cdot  \nabla \delta u_2
	+P\begin{pmatrix}
		u_{ 1y}u_{ 2z}-u_{ 2y}u_{ 1z} \\ 
		u_{ 2x}u_{ 1z}-u_{ 1x}u_{ 2z} \\ 
		u_{ 1x}u_{ 2y}-u_{ 2x}u_{ 1y} \notag   
	\end{pmatrix}\cdot\nabla \delta u_3\\
	+P&(\delta u_{1x}u_{2y}+ u_{1x}\delta u_{2y}+ \delta u_{1x}u_{3z}+ u_{1x}\delta u_{3z}+ \delta u_{2y}u_{3z}+  u_{2y}\delta u_{3z}\\
	-& \delta u_{1y}u_{2x}- u_{1y}\delta u_{2x}- \delta u_{1z}u_{3x}-  u_{1z}\delta u_{3x}- \delta u_{2z}u_{3y}- u_{2z}\delta u_{3y})]d\pmb{\omega}\\
	=&\int_{\mathrm{\Omega}} [P\nabla\cdot\delta\pmb{u}+P\begin{pmatrix}
		u_{ 2y}u_{ 3z}-u_{ 3y}u_{ 2z} \\ 
		u_{ 3x}u_{ 2z}-u_{ 2x}u_{ 3z} \\ 
		u_{ 2x}u_{ 3y}-u_{ 2y}u_{ 3x} \notag   
	\end{pmatrix}\cdot  \nabla \delta u_1\\
	+P&\begin{pmatrix}
		u_{ 3y}u_{ 1z}-u_{ 1y}u_{ 3z} \\ 
		u_{ 1x}u_{ 3z}-u_{ 1z}u_{ 3x} \\ 
		u_{ 3x}u_{ 1y}-u_{ 1x}u_{ 3y} \notag   
	\end{pmatrix}\cdot  \nabla \delta u_2
	+P\begin{pmatrix}
		u_{ 1y}u_{ 2z}-u_{ 2y}u_{ 1z} \\ 
		u_{ 2x}u_{ 1z}-u_{ 1x}u_{ 2z} \\ 
		u_{ 1x}u_{ 2y}-u_{ 2x}u_{ 1y} \notag   
	\end{pmatrix}\cdot\nabla \delta u_3\\
	+P&\begin{pmatrix}
		u_{ 3z}+u_{ 2z} \\ 
		-u_{ 2x} \\ 
		-u_{ 3x} \notag   
	\end{pmatrix}\cdot\nabla \delta u_1+P\begin{pmatrix}
	-u_{ 1y} \\ 
	u_{ 3z}+u_{ 1x} \\ 
	-u_{ 3y} \notag   
	\end{pmatrix}\cdot\nabla \delta u_2+P\begin{pmatrix}
	-u_{ 1z} \\ 
	-u_{ 2z} \\ 
	u_{ 2y}+u_{ 1x} \notag   
	\end{pmatrix}\cdot\nabla \delta u_3]d\pmb{\omega}\\
	\end{aligned}
\end{equation*}

Here, the ``big vectors" are now denoted as $a_{i}$ and $b_{i}$, $i=1,2,3$. By $Green$'s identities with fixed boundary condition, it can be rewrote as   
\begin{equation*}
	\begin{aligned}	
		\delta L_{1}(\pmb{u})&=\int_{\mathrm{\Omega}} [-\nabla P\cdot\delta\pmb{u}-\nabla\cdot P\begin{pmatrix}
	u_{ 2y}u_{ 3z}-u_{ 3y}u_{ 2z} \\ 
	u_{ 3x}u_{ 2z}-u_{ 2x}u_{ 3z} \\ 
	u_{ 2x}u_{ 3y}-u_{ 2y}u_{ 3x} \notag   
\end{pmatrix} \delta u_1\\
-\nabla\cdot P&\begin{pmatrix}
	u_{ 3y}u_{ 1z}-u_{ 1y}u_{ 3z} \\ 
	u_{ 1x}u_{ 3z}-u_{ 1z}u_{ 3x} \\ 
	u_{ 3x}u_{ 1y}-u_{ 1x}u_{ 3y} \notag   
\end{pmatrix}\delta u_2
-\nabla\cdot P\begin{pmatrix}
	u_{ 1y}u_{ 2z}-u_{ 2y}u_{ 1z} \\ 
	u_{ 2x}u_{ 1z}-u_{ 1x}u_{ 2z} \\ 
	u_{ 1x}u_{ 2y}-u_{ 2x}u_{ 1y} \notag   
\end{pmatrix} \delta u_3\\
-\nabla\cdot P&\begin{pmatrix}
	u_{ 3z}+u_{ 2z} \\ 
	-u_{ 2x} \\ 
	-u_{ 3x} \notag   
\end{pmatrix}\nabla \delta u_1-\nabla\cdot P\begin{pmatrix}
	-u_{ 1y} \\ 
	u_{ 3z}+u_{ 1x} \\ 
	-u_{ 3y} \notag   
\end{pmatrix}\nabla \delta u_2-\nabla\cdot P\begin{pmatrix}
	-u_{ 1z} \\ 
	-u_{ 2z} \\ 
	u_{ 2y}+u_{ 1x} \notag   
\end{pmatrix}\delta u_3]d\pmb{\omega}\\
=\int_{\mathrm{\Omega}}& [-\nabla P\cdot\delta\pmb{u}-a_1 \delta u_1
-a_2\delta u_2
-a_3 \delta u_3
-b_1\delta u_1-b_2\delta u_2-b_3\delta u_3]d\pmb{\omega}\\
=\int_{\mathrm{\Omega}}& [-\nabla P\cdot\delta\pmb{u}-\pmb{a} \cdot\delta\pmb{u} -\pmb{b} \cdot\delta \pmb{u}]d\pmb{\omega}=\int_{\mathrm{\Omega}} [-(\nabla P+\pmb{a}  +\pmb{b} )\cdot\delta \pmb{u}]d\pmb{\omega}.\\
	\end{aligned}
\end{equation*}
For some $A_{i}$ such that $\mathrm{\Delta} A_{i}=-((\nabla P)_{i}+a_{i}  +b_{i} )$, $i=1,2,3$, then it goes,
\begin{equation}\label{loss1delF}
		\delta L_{1}(\pmb{u})=\int_{\mathrm{\Omega}} [-(\nabla P+\pmb{a}  +\pmb{b} )\cdot\delta \pmb{u}]d\pmb{\omega}=\int_{\mathrm{\Omega}}[ \mathrm{\Delta}\pmb{A}\cdot\delta \pmb{u}]d\pmb{\omega}=\int_{\mathrm{\Omega}} [\pmb{A}\cdot\delta \mathrm{\Delta}\pmb{u}]d\pmb{\omega}\\
\end{equation}
Next, for the part $L_{2}(\pmb{u})$, one derive
\begin{equation*}
	\begin{aligned}
	\delta L_{2}(\pmb{u})&=\frac{1}{2}\int_{\mathrm{\Omega}}\delta|\nabla \times\pmb{u}|^2]d\pmb{\omega}=\int_{\mathrm{\Omega}} [\begin{pmatrix}
		Q_1 \\ 
		Q_2 \\ 
		Q_3 \notag   
	\end{pmatrix}
	\cdot\begin{pmatrix} \delta u_{3y}- \delta u_{2z}\\
		-\delta u_{1z} + \delta u_{3x}\\
		 \delta u_{2x}- \delta u_{1y}\end{pmatrix}]d\pmb{\omega}\\
	 =&\int_{\mathrm{\Omega}} [\begin{pmatrix}
	 	0 \\ 
	 	-Q_3 \\ 
	 	Q_2 \notag   
	 \end{pmatrix}\cdot  \nabla \delta u_{1}+\begin{pmatrix}
	 	Q_3 \\ 
	 	0 \\ 
	 	-Q_1 \notag   
	 \end{pmatrix}\cdot  \nabla \delta u_{2}
	 +\begin{pmatrix}
	 	-Q_2 \\ 
	 	Q_1 \\ 
	 	0 \notag
	 \end{pmatrix}\cdot  \nabla \delta u_{3}]d\pmb{\omega}\\
 =&\int_{\mathrm{\Omega}} [-\nabla\cdot\begin{pmatrix}
 	0 \\ 
 	-Q_3 \\ 
 	Q_2 \notag   
 \end{pmatrix}  \delta u_{1}-\nabla\cdot\begin{pmatrix}
 	Q_3 \\ 
 	0 \\ 
 	-Q_1 \notag   
 \end{pmatrix}\delta u_{2}
 -\nabla\cdot\begin{pmatrix}
 	-Q_2 \\ 
 	Q_1 \\ 
 	0 \notag
 \end{pmatrix} \delta u_{3}]d\pmb{\omega}\\
	\end{aligned}
\end{equation*}

Denote the ``tall vectors" as $c_{i}$. For some $B_{i}$ such that $\mathrm{\Delta} B_{i}=c_{i}$, $i=1,2,3$, so
\begin{equation}\label{loss2delF}
	\begin{aligned}
		\delta L_{2}
		&=\int_{\mathrm{\Omega}} [\mathrm{\Delta}\pmb{B} \cdot \delta \pmb{u}]d\pmb{\omega}=\int_{\mathrm{\Omega}} [\pmb{B} \cdot \delta \mathrm{\Delta}\pmb{u}]d\pmb{\omega}.
	\end{aligned}
\end{equation}
The constraint (\ref{ssdconst}) implies $\mathrm{\Delta} \pmb{\phi}=\mathrm{\Delta} \pmb{u}=\pmb{F}$ and $\delta\mathrm{\Delta} \pmb{u}=\delta\pmb{F}$. Finally, we have
\begin{equation}\label{lossdelF}
	\begin{aligned}
	\delta L(\pmb{u})&=	\delta L_{1}(\pmb{u})+	\delta L_{2}(\pmb{u})
	=\int_{\mathrm{\Omega}} [\pmb{A}\cdot\delta \mathrm{\Delta}\pmb{u}]d\pmb{\omega}+\int_{\mathrm{\Omega}} [\pmb{B} \cdot \delta \mathrm{\Delta}\pmb{u}] d\pmb{\omega}\\
	=\int_{\mathrm{\Omega}} [\pmb{A}\cdot&\delta \pmb{F}] d\pmb{\omega}+\int_{\mathrm{\Omega}} [ \pmb{B}\cdot \delta \pmb{F}] d\pmb{\omega} \quad\Rightarrow \quad \frac{\partial L}{\partial \pmb{F}}=	\frac{\partial L_{1}}{\partial \pmb{F}}+	\frac{\partial L_{2}}{\partial \pmb{F}}=\pmb{A}+\pmb{B}.
\end{aligned}
\end{equation}
where $\mathrm{\Delta} \pmb{A}=-(\nabla P+\pmb{a} +\pmb{b})$ and $\mathrm{\Delta} \pmb{B}=-\pmb{c}$. 
\begin{remark}
	A similar derivation to (\ref{loss1delF}), (\ref{loss2delF}) and  (\ref{lossdelF}) also works for the case of 2D, only differed by there is no need to treat the Tail$(\pmb{u})$ term in 2D scenario.
\end{remark}

\newpage
\subsection*{A gradient-based numerical scheme based on above derivation}
The algorithm goes as,
\begin{algorithm}[H]
	{
		\caption{$[\pmb{u}]  =  $ shrinkJDandCurl($\pmb{\phi}_{o}=\pmb{id}+\pmb{u}_{o}$)}\label{alg}
		\hrule
		\begin{itemize}
			\item[$\bullet$] 1: initialize $\pmb{F}=\pmb{0}$,  $\pmb{u}=\pmb{u}_{o}$, $\triangle t$ and $ratio$-tolerance;	
			\item[$\bullet$] 2: while $ratio\ge ratio$-tolerance
			\begin{itemize}
				\item[$\bullet$] 3: if $better$
				\begin{itemize}
					\item[$\bullet$] 4: find $\pmb{A}$ and $\pmb{B}$ from $\mathrm{\Delta} \pmb{A}=-(\nabla P+\pmb{a} +\pmb{b})$ and $\mathrm{\Delta} \pmb{B}=-\pmb{c}$;		
				\end{itemize}
				\item[$\bullet$] 5: update $\pmb{F}_{new}=\pmb{F}-\triangle t * (\pmb{A}+\pmb{B})$;
				\item[$\bullet$] 6: solve for $\pmb{u}$ from		$\mathrm{\Delta}\pmb{u}=\pmb{F}$;
				\item[$\bullet$] 7: if $L$ decrease,
				\begin{itemize}
					\item[$\bullet$] 9: $better$; 
					\item[$\bullet$] 10: $\triangle t=\triangle t * t_{up}$ (e.e, set $t_{up}=1.01$ and $t_{down}=0.99$);
					\item[$\bullet$] 11: $\pmb{F}=\pmb{F}_{new}$;
				\end{itemize}
				\hspace{0.3cm} else
				\begin{itemize}
					\item[$\bullet$] 12: $better$ false;
					\item[$\bullet$] 13: $\triangle t=\triangle t * t_{down}$.
				\end{itemize}
			\end{itemize}
		\end{itemize}	
	}
\end{algorithm}

\end{document}